\documentclass[showkeys,superscriptaddress,floatfix,prd,10pt,aps]{revtex4-2}
\usepackage{graphicx,epstopdf}
\usepackage[caption=false]{subfig}
\usepackage{appendix}
\usepackage[T1]{fontenc}
\usepackage{lmodern}
\usepackage[dvipsnames,x11names]{xcolor}
\usepackage[colorlinks=true,linkcolor=NavyBlue,citecolor=ForestGreen,urlcolor=NavyBlue]{hyperref}
\usepackage[sort&compress]{natbib}
\usepackage{lipsum}
\usepackage{morefloats}
\usepackage[pdf]{pstricks}
\usepackage{amsmath}
\usepackage{amssymb}
\usepackage{amsfonts}
\usepackage{rotating}
\usepackage{cancel}
\usepackage{mathtools}
\usepackage{bbm}
\usepackage{dsfont}
\usepackage{bbold}
\usepackage{multirow}
\usepackage{ulem}
\usepackage{physics}
\usepackage{orcidlink}
\usepackage{colortbl}

\def\pslash{p\!\!\!\slash }
\def\qslash{q\!\!\!\slash }
\def\xslash{x\!\!\!\slash }
\def\eslash{\varepsilon\!\!\!\slash }

\begin{document}

\title{Electromagnetic properties of $\bar D^{(*)}\Xi^{\prime}_c$, $\bar D^{(*)}\Lambda_c$,  $\bar D_s^{(*)}\Lambda_c$ and  $\bar D_s^{(*)}\Xi_c$  pentaquarks}

\author{Ula\c{s} \"{O}zdem\orcidlink{0000-0002-1907-2894}}%
\email[]{ulasozdem@aydin.edu.tr}
\affiliation{Health Services Vocational School of Higher Education, Istanbul Aydin University, Sefakoy-Kucukcekmece, 34295 Istanbul, T\"{u}rkiye}

\date{\today}
 
\begin{abstract}
To elucidate the internal structure of exotic states is one of the central purposes of hadron physics.  Motivated by this, we study the electromagnetic properties of $\bar D^{(*)}\Xi^{\prime}_c$, $\bar D^{(*)}\Lambda_c$,  $\bar D_s^{(*)}\Lambda_c$ and  $\bar D_s^{(*)}\Xi_c$  pentaquarks without strange, with strange and with double strange through QCD light-cone sum rules. We have also evaluated electric quadrupole and magnetic octupole moments of the $\bar D^{*}\Xi^{\prime}_c$, $\bar D^{*}\Lambda_c$,  $\bar D_s^{*}\Lambda_c$ and  $\bar D_s^{*}\Xi_c$ pentaquarks.   The magnetic dipole moment is the leading-order response of a bound system to a soft external magnetic field.  Thus, it ensures a prominent platform for the examination of the internal organizations of hadrons governed by the quark-gluon dynamics of QCD. We look forward to the present study stimulating the interest of experimentalists in investigating the electromagnetic properties of the hidden-charm pentaquarks.
\end{abstract}

\maketitle

\section{Introduction}\label{motivation}

Although it was suggested long ago that non-conventional states other than standard hadrons might exist, the experimental evidence of these states was taken to a new level in 2003 with the Belle Collaboration observation of the X(3872) particle \cite{Belle:2003nnu}.
Since then a series of states beyond the non-conventional states have been discovered. Most of those are potentially exotic states such as tetraquarks, pentaquarks, hybrid mesons, glueballs, and so on. The exploration of exotic states, particularly how the quarks are grouped inside (i.e., molecular or compact configuration) plays an important role in the comprehension of the behavior of the low-energy QCD. Therefore, investigating the properties of these states and shedding light on their nature is one of the most active areas of research in both experimental and theoretical hadron physics \cite{Esposito:2014rxa,Esposito:2016noz,Olsen:2017bmm,Lebed:2016hpi,Nielsen:2009uh,Brambilla:2019esw,Agaev:2020zad,Chen:2016qju,Ali:2017jda,Guo:2017jvc,Liu:2019zoy,Yang:2020atz,Dong:2021juy,Dong:2021bvy,Meng:2022ozq, Chen:2022asf}.

In 2015, the LHCb Collaboration reported two resonance states, $P_{c }(4380)$ and $P_{c }(4450)$, in the  $J/\psi\,p$ invariant mass distribution of the $\Lambda_b \rightarrow J/\psi\,p\,K$ decay~\cite{Aaij:2015tga}. In 2019, the LHCb Collaboration updated the observations in the  $\Lambda_b \rightarrow J/\psi\,p\,K$  process by using more data \cite{Aaij:2019vzc}, they not only discovered a new narrow pentaquark state, $P_{c}(4312)$, but also found that the $P_{c }(4450)$  consists of two narrow overlapping peaks, $P_{c}(4440)$, and $P_{c}(4457)$.  In 2020, the LHCb collaboration reported evidence of a hidden-charm pentaquark candidate, $P_{cs}(4459)$, in the invariant mass spectrum of $J/\psi\Lambda$ in the $\Xi_b^0 \rightarrow J/\psi\,\Lambda\,K^-$ decay~\cite{Aaij:2020gdg}.  In 2021, the LHCb Collaboration reported evidence for a new pentaquark candidate, $P_c(4337)$,  in $B_s \rightarrow J/\psi\, p \bar p$ decays~\cite{LHCb:2021chn}. Recently, the LHCb Collaboration announced the
observed a new pentaquark state, $P_{cs}(4338)$, in the $J/\psi\Lambda$ mass distribution in the $B^- \rightarrow J/\psi\Lambda^- p $ decays~\cite{Collaboration:2022boa}. The masses, widths, and minimal valence quark contents of hidden-charm pentaquarks are presented in Table \ref{pentaquarks}. Systematical research on their nature and inner structure can progress our comprehension of the non-perturbative behaviors of strong interaction. 
  Therefore, a great deal of theoretical effort has been devoted to figuring out the nature of the hidden-charm pentaquarks~\cite{Shen:2020gpw, Wang:2019nvm, Wu:2010jy, Chen:2016ryt, Shen:2019evi, Xiao:2019gjd, Anisovich:2015zqa, Feijoo:2015kts, Lu:2016roh, Liu:2020hcv, Zou:2021sha, Karliner:2021xnq, Peng:2020hql, Zhu:2021lhd, Hu:2021nvs, Du:2021bgb,  Xiao:2021rgp, Chen:2020uif, Chen:2015moa, Chen:2016otp, Xiang:2017byz, Chen:2019bip, Chen:2020pac, Chen:2020opr, Chen:2020kco, Chen:2021tip, Chen:2022onm, Chen:2019asm, Wu:2021caw, Lu:2021irg, Yang:2021pio, Cheng:2021gca, Clymton:2021thh, Liu:2020ajv, Deng:2022vkv, Shi:2021wyt, Wang:2022gfb, Wang:2021itn, Wang:2020eep, Azizi:2021utt, Wang:2022neq,Karliner:2022erb, Wang:2022mxy,Yan:2022wuz,Meng:2022wgl,Burns:2022uha, Feijoo:2022rxf, Zhu:2022wpi, Azizi:2022qll, Ortega:2022uyu}.

\begin{table}[htp]
\caption{The masses, widths, and minimal valence quark contents of hidden-charm pentaquarks.}\label{pentaquarks}
\begin{tabular}{l|c|c|cc}
\toprule
State  & Mass (MeV) & Width (MeV) & Content \\
\toprule
$P_c(4380)^+$  &            $4380\pm8\pm29$          &       $215\pm18\pm86$           & $uudc\bar{c}$ \\
$P_c(4312)^+$       & ~~$4311.9\pm0.7^{~+6.8}_{~-0.6}$~~  &  $9.8\pm2.7^{~+3.7}_{~-4.5}$    & $uudc\bar{c}$ \\
$P_c(4440)^+$      &    $4440.3\pm1.3^{~+4.1}_{~-4.7}$   &  $20.6\pm4.9^{~+8.7}_{~-10.1}$  & $uudc\bar{c}$ \\
$P_c(4457)^+$      &    $4457.3\pm0.6^{~+4.1}_{~-1.7}$   &  $6.4\pm2.0^{~+5.7}_{~-1.9}$    & $uudc\bar{c}$ \\
$P_{cs}(4459)^0$   &    $4458.8\pm2.9^{~+4.7}_{~-1.1}$   &  $17.3\pm6.5^{~+8.0}_{~-5.7}$   & $udsc\bar{c}$ \\
$P_c(4337)^+$       &       $4337^{~+7~+2}_{~-4~-2}$      &  $29^{~+26~+14}_{~-12~-14}$     & $uudc\bar{c}$ \\
$P_{cs}(4338)^0$   &    $4338.2 \pm 0.7 \pm 0.4$   &  $7.0 \pm 1.2 \pm 1.3$    & $udsc\bar{c}$ \\
\toprule
\end{tabular}
\end{table}

The discovery of pentaquarks in the $J/\psi\,p$ system at LHCb opened a new era of examination in hadron spectroscopy. Although mass spectra, decay behavior, and production mechanisms of pentaquarks have attracted much attention on both the theoretical and experimental sides, examination of the electromagnetic features of these states has not received plenty of attention ~\cite{Wang:2016dzu, Ozdem:2018qeh, Ortiz-Pacheco:2018ccl,Xu:2020flp,Ozdem:2021btf, Li:2021ryu,Ozdem:2021ugy,Gao:2021hmv, Ozdem:2022iqk, Ozdem:2022vip, Ozdem:2022kei, Wang:2022tib}. The magnetic dipole moment is another intrinsic observable of hadrons which may contain prominent data of its quark-gluon structure and underlying dynamics.  Different magnetic dipole moments will affect both the total and differential cross sections in the photo- or electro-production of pentaquarks. Therefore, such an investigation will deepen our knowledge of pentaquarks and help us understand the underlying dynamics governing their formation. 
As expected, since the masses of these states are slightly below or above the thresholds of the charmed meson-baryon pairs, the  $P_c(4312)$, $P_c(4380)$, $P_c(4440)$, $P_c(4457)$, $P_{cs}(4459)$ and $P_{cs}(4338)$ have been assigned to be the $\bar{D}\Sigma_c$, $\bar{D}\Sigma_c^*$, $\bar{D}^*\Sigma_c$, $\bar{D}^*\Sigma_c^*$, $\bar{D}^*\Xi_c$, and $\bar{D}\Xi_c$, respectively, have been found  to be the charmed meson-baryon molecular states according to several phenomenological analyses. Following the experimental observations of these hidden-charm pentaquarks, numerous studies have begun to investigate other possible hidden-charm pentaquarks and their properties.  Inspired by these studies and our previous work, we focused on two possible questions: Can we calculate the magnetic dipole moments of other possible hidden-charm pentaquarks in the molecular structure? And what are the electromagnetic properties of other possible molecular hidden-charm pentaquarks with and without strangeness? Motivated by these questions, we investigate the magnetic dipole moment of possible  $\bar D^{(*)}\Xi^{\prime}_c$, $\bar D^{(*)}\Lambda_c$,  $\bar D_s^{(*)}\Lambda_c$, and  $\bar D_s^{(*)}\Xi_c$ hidden-charm pentaquarks without strange, with strange and with double strange. We would like to point out that there are other possible hidden-charm pentaquarks besides the ones we have examined in this study, but we cannot calculate the magnetic dipole moments of these possible hidden-charm pentaquarks because both the mass and the residue values have not yet been calculated. If these values are calculated in the future, we may be able to calculate the magnetic dipole moments of other possible hidden-charm pentaquarks.
Since the magnetic dipole moments belong to the non-perturbative domain of the QCD, we need non-perturbative methods to be able to perform the calculations of these parameters. The QCD light-cone sum rule is one of the powerful methods developed to calculate non-perturbative effects. In this technique, a proper correlation function is computed concerning hadronic parameters and their magnetic dipole moment on one side and quark-gluon degrees of freedom and distribution amplitudes of the on-shell photon on the other side. The distribution amplitudes of photon are expressed concerning functions of different twists. Then, continuum subtraction and Borel transform procedures provided by the quark-hadron duality ansatz are carried out to remove the contributions coming from the higher states and continuum. By equating the results acquired in two different regions, the related sum rules for the magnetic dipole moment of the hadrons are obtained~\cite{Chernyak:1990ag, Braun:1988qv, Balitsky:1989ry}.  

This paper is organized as follows.  We introduce our theoretical framework in Sec. \ref{formalism} and the corresponding numerical results and discussions in Sec. \ref{numerical}.  Finally, a summary is given in Sec. \ref{summary}. The analytical expressions acquired for the magnetic dipole moment of the $\bar D \Xi^{\prime}_c$ pentaquark state are given in the Appendix.

\begin{widetext}
 
\section{The QCD light-cone sum rules for hidden-charm pentaquarks}\label{formalism}

\subsection{Formalism of the  spin-$\frac{1}{2}$ pentaquarks} 

The correlation function required to perform magnetic dipole moment analysis of the spin-1/2 pentaquarks ( hereafter $P_{c \bar c}^{1/2} $ ) is given in the following form:
\begin{eqnarray} \label{edmn01}
\Pi(p,q)&=&i\int d^4x e^{ip \cdot x} \langle0|T\left\{J^{P_{c\bar c}^{1/2}}(x)\bar{J}^{P_{c\bar c}^{1/2}}(0)\right\}|0\rangle _\gamma \, , 
\end{eqnarray}
where  $q$ is the momentum of the photon, sub-indice $\gamma$ is the external electromagnetic field,   and $J(x)$ stand for the interpolating currents of  the $P_{c\bar c}^{1/2}$ states.  The corresponding interpolating currents are given by
\begin{align}\label{curpcs1}
J^{1}(x)&=\frac{1}{\sqrt{2}}\Big \{ \mid \bar D^0 \Xi^{\prime 0}_c \rangle \, - \mid \bar D^- \Xi^{\prime +}_c \rangle  \Big \}=\frac{1}{\sqrt{2}} \Big \{ \big[\bar c^d(x)i \gamma_5 u^d(x)\big]\big[\varepsilon^{abc} d^{a^T}(x)C\gamma_\mu s^b(x) \gamma^\mu \gamma_5 c^c(x)\big]  \nonumber\\
&
- \big[\bar c^d(x)i \gamma_5 d^d(x)\big]\big[\varepsilon^{abc} u^{a^T}(x)C\gamma_\mu s^b(x)\gamma^\mu \gamma_5
 c^c(x)\big] \Big\} \, ,\nonumber\\
 J^{2}(x)&=\mid \bar D^0 \Lambda^{+}_c \rangle   = \Big \{ \big[\bar c^d(x) i\gamma_5 u^d(x)\big]
 \big[\varepsilon^{abc}u^{a^T}(x)C\gamma_5 d^b(x) c^c(x)\big]\Big \}\, ,  \nonumber\\
J^{3}(x)&=\mid \bar D^- \Lambda^{+}_c \rangle   = \Big \{ \big[\bar c^d(x) i\gamma_5 d^d(x)\big]
 \big[\varepsilon^{abc}u^{a^T}(x)C\gamma_5 d^b(x) c^c(x)\big]\Big \}\, ,  \nonumber\\
J^{4}(x)&=\mid  \bar D_s^- \Xi^{0}_c \rangle   = \Big \{ \big[\bar c^d(x) i\gamma_5 s^d(x)\big]
 \big[\varepsilon^{abc}d^{a^T}(x)C\gamma_5 s^b(x) c^c(x)\big]\Big \}\, ,  \nonumber\\
J^{5}(x)&=\mid \bar D_s^- \Lambda^{+}_c \rangle   = \Big \{ \big[\bar c^d(x) i\gamma_5 s^d(x)\big]
 \big[\varepsilon^{abc}u^{a^T}(x)C\gamma_5 d^b(x) c^c(x)\big]\Big \}\, ,  
 \end{align}
where $u(x)$, $d(x)$, $s(x)$ and $c(x)$ being quark fields, $a$, $b$, $c$ and  $d$ stand for color indices; and the $C$ denotes the charge conjugation operator.

On the QCD side, we contract the relevant quark operators in the correlation function at the quark-gluon level with the help of Wick's theorem. After these manipulations, the correlation function of the QCD side is obtained in terms of the light and heavy-quark propagators, and distribution amplitudes (DAs) of the photon.  
For example, in the case of the  state $\bar D \Xi^{\prime}_c$, the result of the contractions is obtained as follows
\begin{align}
\label{QCD1}
\Pi_1^{QCD}(p,q)&=- \frac{i}{2}\varepsilon^{abc} \varepsilon^{a^{\prime}b^{\prime}c^{\prime}}\, \int d^4x \, e^{ip\cdot x} \langle 0\mid \Big\{ 
\, \mbox{Tr}\Big[\gamma_5 S_{u}^{dd^\prime}(x) \gamma_5  S_{c}^{d^\prime d}(-x)\Big]  
\mbox{Tr}\Big[\gamma_\mu S_s^{bb^\prime}(x) \gamma_\nu \widetilde S_{d}^{aa^\prime}(x)\Big]
\nonumber\\
&
- \mbox{Tr}\Big[\gamma_5 S_{u}^{da^\prime}(x) \gamma_\nu  \widetilde S_s^{bb^\prime}(x) \gamma_\mu S_{d}^{ad^\prime}(x)   \gamma_5 S_{c}^{d^\prime d}(-x)\Big] 
\nonumber\\ 
&
 - \mbox{Tr}\Big[\gamma_5 S_{d}^{da^\prime}(x) \gamma_\nu  \widetilde S_s^{bb^\prime}(x) \gamma_\mu  S_{u}^{ad^\prime}(x) \gamma_5 S_{c}^{d^\prime d}(-x)\Big]
     \nonumber\\ 
&   +
 \mbox{Tr}\Big[\gamma_5     S_{d}^{dd^\prime}(x) \gamma_5 S_{c}^{d^\prime d}(-x)\Big]  
\mbox{Tr}\Big[\gamma_\mu S_s^{bb^\prime}(x)
  \gamma_\nu 
    \widetilde S_{u}^{aa^\prime}(x)\Big] \Big\} \Big( \gamma_\mu \gamma_5 S_c^{cc^{\prime}}(x)\gamma_5 \gamma_\nu \Big)
\mid 0 \rangle _\gamma \, , 
\end{align}
where   
$\widetilde{S}_{c(q)}^{ij}(x)=CS_{c(q)}^{ij\mathrm{T}}(x)C$ and,
 $S_{c}(x)$ and $S_{q}(x)$ denote the charm and light-quark propagators. The explicit expressions of
these terms can be written as~\cite{Yang:1993bp, Belyaev:1985wza}
\begin{align}
\label{edmn13}
S_{q}(x)&= \frac{1}{2 \pi x^2}\Big(i \frac{\xslash}{x^2}- \frac{m_q}{2}\Big) 
- \frac{\langle \bar qq \rangle }{12} \Big(1-i\frac{m_{q} \xslash}{4}   \Big)
- \frac{ \langle \bar qq \rangle }{192}
m_0^2 x^2  \Big(1-i\frac{m_{q} \xslash}{6}   \Big)
-\frac {i g_s }{32 \pi^2 x^2} ~G^{\mu \nu} (x) 
\Big[\rlap/{x} 
\sigma_{\mu \nu} +  \sigma_{\mu \nu} \rlap/{x}
 \Big],
\end{align}%
\begin{align}
\label{edmn14}
S_{c}(x)&=\frac{m_{c}^{2}}{4 \pi^{2}} \Bigg[ \frac{K_{1}\Big(m_{c}\sqrt{-x^{2}}\Big) }{\sqrt{-x^{2}}}
+i\frac{{\xslash}~K_{2}\Big( m_{c}\sqrt{-x^{2}}\Big)}
{(\sqrt{-x^{2}})^{2}}\Bigg]
-\frac{g_{s}m_{c}}{16\pi ^{2}} \int_0^1 dv\, G^{\mu \nu }(vx)\Bigg[ (\sigma _{\mu \nu }{\xslash}
  +{\xslash}\sigma _{\mu \nu }) \frac{K_{1}\Big( m_{c}\sqrt{-x^{2}}\Big) }{\sqrt{-x^{2}}}
   \nonumber\\
  &
+2\sigma_{\mu \nu }K_{0}\Big( m_{c}\sqrt{-x^{2}}\Big)\Bigg],
\end{align}%
where $\langle \bar qq \rangle$ stands for light-quark condensate, $G^{\mu\nu}$ denotes the gluon field strength tensor, $v$ is line variable, and  $K_i$'s are modified Bessel functions of the second kind.   

The correlation function in Eq.~(\ref{QCD1})  receives different contributions: perturbative, i.e., when a photon interacts perturbatively with quark propagators, and nonperturbative, i.e., photon interacts with light quarks at a large distance, contributions. Details of this procedure applied to obtain the expression of perturbative and non-perturbative contributions are given in Refs.~\cite{Ozdem:2022vip,Ozdem:2022eds}. 
With this procedure, the calculation of the QCD side of the correlation function is completed.

To obtain the second representation of the correlation function, the hadronic side, we isolate the ground state contributions from pentaquarks with spin-parity $J^P =1/2^-$, and acquire the hadronic side,
 \begin{align}\label{edmn02}
\Pi^{Had}(p,q)&=\frac{\langle0\mid J^{P_{c \bar c}^{1/2}}(x) \mid
{P_{c \bar c}^{1/2}}(p, s) \rangle}{[p^{2}-m_{P_{c \bar c}^{1/2}}^{2}]} 
\langle {P_{c \bar c}^{1/2}}(p, s)\mid
{P_{c \bar c}^{1/2}}(p+q, s)\rangle_\gamma 
\frac{\langle {P_{c \bar c}^{1/2}}(p+q, s)\mid
\bar J^{P_{c \bar c}^{1/2}}(0) \mid 0\rangle}{[(p+q)^{2}-m_{P_{c \bar c}^{1/2}}^{2}]}+ \cdots 
\end{align}
where dots represent the contributions of excited states and continuum.

The matrix elements in the above equation are given in terms of hadronic parameters and form factors: 
%
\begin{align} 
\langle0\mid J^{P_{c \bar c}^{1/2}}(x)\mid {P_{c \bar c}^{1/2}}(p, s)\rangle=&\lambda_{P_{c \bar c}^{1/2}} \gamma_5 \, u(p,s),\label{edmn04}\\
\nonumber\\
\langle {P_{c \bar c}^{1/2}}(p+q, s)\mid\bar J^{P_{c \bar c}^{1/2}}(0)\mid 0\rangle=&\lambda_{P_{c \bar c}^{1/2}} \gamma_5 \, \bar u(p+q,s)\label{edmn004}
,\\
\nonumber\\
\langle {P_{c \bar c}^{1/2}}(p, s)\mid {P_{c \bar c}^{1/2}}(p+q, s)\rangle_\gamma &=\varepsilon^\mu\,\bar u(p, s)\Big[\big[F_1(q^2)
+F_2(q^2)\big] \gamma_\mu +F_2(q^2)
\frac{(2p+q)_\mu}{2 m_{P_{c \bar c}^{1/2}}}\Big]\,u(p+q, s). \label{edmn005}
\end{align}

Then Eqs. (\ref{edmn04})-(\ref{edmn005}) are inserted in the Eq. (\ref{edmn02}), we get the following result for the hadronic side of the correlation function,
\begin{align}
\label{edmn05}
\Pi^{Had}(p,q)=&\lambda^2_{P_{c \bar c}^{1/2}}\gamma_5 \frac{\Big(\pslash+m_{P_{c \bar c}^{1/2}} \Big)}{[p^{2}-m_{{P_{c \bar c}^{1/2}}}^{2}]}\varepsilon^\mu \Bigg[\big[F_1(q^2) %
+F_2(q^2)\big] \gamma_\mu
+F_2(q^2)\, \frac{(2p+q)_\mu}{2 m_{P_{c \bar c}^{1/2}}}\Bigg]  \gamma_5 
\frac{\Big(\pslash+\qslash+m_{P_{c \bar c}^{1/2}}\Big)}{[(p+q)^{2}-m_{{P_{c \bar c}^{1/2}}}^{2}]}. 
\end{align}
In obtaining the above analytical expression, summation over spins of $P_{c\bar c}^{1/2}$
\begin{align}
\label{edmn0004}
 \sum_s u(p,s)\bar u(p,s)&=\pslash+m_{P_{c\bar c}^{1/2}},\\
  \sum_s u(p+q,s)\bar u(p+q,s)&=(\pslash+\qslash)+m_{P_{c\bar c}^{1/2}},
\end{align}
were also carried out. 

The  magnetic dipole form factor, $F_M(q^2)$, is written with respect to the $F_1(q^2)$ and $F_2(q^2)$ form factor at different $q^2$ :
\begin{align}
\label{edmn07}
&F_M(q^2) = F_1(q^2) + F_2(q^2).
\end{align}
 In case of on-shell photon, i.e. $q^2 = 0 $,  the 
 the magnetic dipole form factor is proportional to the magnetic dipole moment $\mu_{P_{c \bar c}^{1/2}}$:
\begin{align}
\label{edmn08}
&\mu_{P_{c \bar c}^{1/2}} = \frac{ e}{2\, m_{P_{c \bar c}^{1/2}}} \,F_M(q^2 = 0).
\end{align}

The QCD light-cone sum rules for the magnetic dipole moments of the spin-1/2 pentaquarks are extracted by equating the coefficients of
the structure $\eslash \qslash$ from the hadronic and QCD sides of the correlation function. To remove the effects coming from the higher states and continuum, Borel transformation as well as continuum subtraction provided by the quark-hadron duality approximation are used.  When all the above-mentioned procedures are performed, the following results are obtained for the magnetic dipole moments:
\begin{align}
\label{edmn15}
\mu_1 \,\lambda^2_1\, m_1 &=e^{\frac{m^2_1}{M^2}}\, \Delta_1^{QCD} (M^2,s_0),~~~~~~~~
\mu_2 \,\lambda^2_2\, m_2 =e^{\frac{m^2_2}{M^2}}\, \Delta_2^{QCD} (M^2,s_0),\nonumber\\
\mu_3 \,\lambda^2_3\, m_3 &=e^{\frac{m^2_3}{M^2}}\, \Delta_3^{QCD} (M^2,s_0),~~~~~~~~
\mu_4 \,\lambda^2_4\, m_4 =e^{\frac{m^2_4}{M^2}}\, \Delta_4^{QCD} (M^2,s_0),\nonumber\\
\mu_5 \,\lambda^2_5\, m_5 &=e^{\frac{m^2_5}{M^2}}\, \Delta_5^{QCD} (M^2,s_0),
\end{align}
where $\mu_i$, $m_i$, and $\lambda_i$ are magnetic dipole moment, mass, and residue of the related pentaquarks. 
As an example, only the explicit expressions of the $\Delta_1^{QCD} (M^2,s_0)$ function are given in the appendix, the rest of the  $\Delta_i^{QCD} (M^2,s_0)$ expressions are in similar forms.

\subsection{Formalism of the  spin-$\frac{3}{2}$ pentaquarks} 

For the magnetic dipole moment of spin-3/2 pentaquarks (here after $P_{c \bar c}^{3/2} $) needed correlation function is introduced as 
\begin{eqnarray} \label{Pc101}
\Pi_{\mu\nu}(p,q)&=&i\int d^4x e^{ip \cdot x} \langle0|T\left\{J_\mu^{P_{c \bar c}^{3/2}}(x)\bar{J}_\nu^{P_{c \bar c}^{3/2}}(0)\right\}|0\rangle _\gamma \, .
\end{eqnarray}
The interpolating currents used in the magnetic dipole moment analyses of $P_{c \bar c}^{3/2} $ pentaquarks are written as follows:
\begin{align}\label{curpcs2}
J_\mu^{1}(x)&=\frac{1}{\sqrt{2}}\Big \{ \mid \bar D^{*0} \Xi^{\prime 0}_c \rangle \, - \mid \bar D^{*-} \Xi^{\prime +}_c \rangle  \Big \}=\frac{1}{\sqrt{2}} \Big \{ \big[\bar c^d(x)\gamma_\mu u^d(x)\big]\big[\varepsilon^{abc} d^{a^T}(x)C\gamma_\alpha s^b(x) \gamma^\alpha \gamma_5 c^c(x)\big]  \nonumber\\
&
- \big[\bar c^d(x)\gamma_\mu d^d(x)\big]\big[\varepsilon^{abc} u^{a^T}(x)C\gamma_\alpha s^b(x)\gamma^\alpha \gamma_5
 c^c(x)\big] \Big\} \, ,\nonumber\\
 J_\mu^{2}(x)&=\mid \bar D^{*0} \Lambda^{+}_c \rangle   = \Big \{ \big[\bar c^d(x) \gamma_\mu u^d(x)\big]
 \big[\varepsilon^{abc}u^{a^T}(x)C\gamma_5 d^b(x) c^c(x)\big]\Big \}\, ,  \nonumber\\
J_\mu^{3}(x)&=\mid \bar D^{*-} \Lambda^{+}_c \rangle   = \Big \{ \big[\bar c^d(x) \gamma_\mu d^d(x)\big]
 \big[\varepsilon^{abc}u^{a^T}(x)C\gamma_5 d^b(x) c^c(x)\big]\Big \}\, ,  \nonumber\\
J_\mu^{4}(x)&=\mid  \bar D_s^{*-} \Xi^{0}_c \rangle   = \Big \{ \big[\bar c^d(x) \gamma_\mu s^d(x)\big]
 \big[\varepsilon^{abc}d^{a^T}(x)C\gamma_5 s^b(x) c^c(x)\big]\Big \}\, ,  \nonumber\\
J_\mu^{5}(x)&=\mid \bar D_s^{*-} \Lambda^{+}_c \rangle   = \Big \{ \big[\bar c^d(x) \gamma_\mu s^d(x)\big]
 \big[\varepsilon^{abc}u^{a^T}(x)C\gamma_5 d^b(x) c^c(x)\big]\Big \}\,.  
 \end{align}

In this part of our analysis, our goal is to get the correlation function with the quark-gluon parameters and photon DAs. 
As an example for the $\bar D^* \Xi^{\prime}_c$ state, if we repeat the steps used in the analysis of spin-1/2 pentaquarks, we obtain the following result:
\begin{align}
\label{QCD6}
\Pi_{\mu\nu-1}^{QCD}(p,q)&= \frac{i}{2}\varepsilon^{abc} \varepsilon^{a^{\prime}b^{\prime}c^{\prime}}\, \int d^4x \, e^{ip\cdot x} \langle 0\mid \Big\{ 
\, \mbox{Tr}\Big[\gamma_\mu S_{u}^{dd^\prime}(x) \gamma_\nu  S_{c}^{d^\prime d}(-x)\Big]  
\mbox{Tr}\Big[\gamma_\alpha S_s^{bb^\prime}(x) \gamma_\beta \widetilde S_{d}^{aa^\prime}(x)\Big]
\nonumber\\
&
- \mbox{Tr}\Big[\gamma_\mu S_{u}^{da^\prime}(x) \gamma_\beta \widetilde S_s^{bb^\prime}(x) \gamma_\alpha S_{d}^{ad^\prime}(x)   \gamma_\nu S_{c}^{d^\prime d}(-x)\Big] 
 - \mbox{Tr}\Big[\gamma_\mu S_{d}^{da^\prime}(x) \gamma_\beta  \widetilde S_s^{bb^\prime}(x) \gamma_\alpha  S_{u}^{ad^\prime}(x) \gamma_\nu S_{c}^{d^\prime d}(-x)\Big]
     \nonumber\\ 
&   +
 \mbox{Tr}\Big[\gamma_\mu     S_{d}^{dd^\prime}(x) \gamma_\nu S_{c}^{d^\prime d}(-x)\Big]  
\mbox{Tr}\Big[\gamma_\alpha S_s^{bb^\prime}(x) \gamma_\beta    \widetilde S_{u}^{aa^\prime}(x)\Big] \Big\} 
    \Big( \gamma^\alpha \gamma_5 S_c^{cc^{\prime}}(x)\gamma_5 \gamma^\beta \Big)
\mid 0 \rangle _\gamma \,. 
\end{align}

As a result of these procedures, the analysis of the QCD side of the correlation function for spin-3/2 pentaquarks is completed.

Our next task is to obtain the relevant correlation function based on hadronic parameters.  It is written as follows,
\begin{align}\label{Pc103}
\Pi^{Had}_{\mu\nu}(p,q)&=\frac{\langle0\mid  J_{\mu}^{P_{c \bar c}^{3/2}}(x)\mid
{P_{c \bar c}^{3/2}}(p,s)\rangle}{[p^{2}-m_{{P_{c \bar c}^{3/2}}}^{2}]}
 \langle {P_{c \bar c}^{3/2}}(p,s)\mid
{P_{c \bar c}^{3/2}}(p+q,s)\rangle_\gamma 
\frac{\langle {P_{c \bar c}^{3/2}}(p+q,s)\mid
\bar{J}_{\nu}^{P_{c \bar c}^{3/2}}(0)\mid 0\rangle}{[(p+q)^{2}-m_{{P_{c \bar c}^{3/2}}}^{2}]}+...
\end{align}
In the above equation, the matrix elements of the interpolating current between  the $P_{c \bar c}^{3/2}$ pentaquark state and the vacuum  are given as:
\begin{align}\label{lambdabey}
\langle0\mid J_{\mu}^{P_{c \bar c}^{3/2}}(x)\mid {P_{c \bar c}^{3/2}}(p,s)\rangle&=\lambda_{{P_{c \bar c}^{3/2}}}u_{\mu}(p,s),\nonumber\\
\langle {P_{c \bar c}^{3/2}}(p+q,s)\mid
\bar{J}_{\nu}^{P_{c \bar c}^{3/2}}(0)\mid 0\rangle &= \lambda_{{P_{c \bar c}^{3/2}}}\bar u_{\nu}(p+q,s), 
\end{align}
where the $\lambda_{{P_{c \bar c}^{3/2}}}$ and $u_{\mu}(p,s)$ ($u_{\nu}(p+q,s)$)  are the residue and   spinors of the $P_{c \bar c}^{3/2}$ pentaquarks, respectively. To further simplify, the summation over the spin of the Rarita-Schwinger spinor for the spin-3/2 pentaquarks is given:
\begin{align}\label{raritabela}
\sum_{s}u_{\mu}(p,s)\bar u_{\nu}(p,s)&=-\Big(\pslash+m_{P_{c \bar c}^{3/2}}\Big)\Big[g_{\mu\nu}
-\frac{1}{3}\gamma_{\mu}\gamma_{\nu}
 -\frac{2\,p_{\mu}p_{\nu}}
{3\,m^{2}_{{P_{c \bar c}^{3/2}}}}+\frac{p_{\mu}\gamma_{\nu}-p_{\nu}\gamma_{\mu}}{3\,m_{{P_{c \bar c}^{3/2}}}}\Big].
\end{align} 
The transition matrix element $\langle
{P_{c \bar c}^{3/2}}(p)\mid {P_{c \bar c}^{3/2}}(p+q)\rangle_\gamma$ entering Eq.
(\ref{Pc103}) can be introduced as 
\cite{Weber:1978dh,Nozawa:1990gt,Pascalutsa:2006up,Ramalho:2009vc}:
\begin{align}\label{matelpar}
\langle {P_{c \bar c}^{3/2}}(p,s)\mid {P_{c \bar c}^{3/2}}(p+q,s)\rangle_\gamma &=-e\bar
u_{\mu}(p,s)\Bigg[F_{1}(q^2)g_{\mu\nu}\eslash 
-
\frac{1}{2m_{{P_{c \bar c}^{3/2}}}} 
\Big[F_{2}(q^2)g_{\mu\nu} 
+F_{4}(q^2)\frac{q_{\mu}q_{\nu}}{(2m_{{P_{c \bar c}^{3/2}}})^2}\Big]\eslash\qslash
\nonumber\\
&+
F_{3}(q^2)\frac{1}{(2m_{{P_{c \bar c}^{3/2}}})^2}q_{\mu}q_{\nu}\eslash \Bigg]
 u_{\nu}(p+q,s).
\end{align}
where $F_i$'s are the Lorentz invariant form factors.  Inserting Eqs. (\ref{lambdabey})-(\ref{matelpar}) into Eq. (\ref{Pc103}) for hadronic side we get 
 %
\begin{align}\label{fizson}
 \Pi^{Had}_{\mu\nu}(p,q)&=-\frac{\lambda_{_{P_{c \bar c}^{3/2}}}^{2}\,\Big(\pslash+m_{P_{c \bar c}^{3/2}}\Big)}{[(p+q)^{2}-m_{_{P_{c \bar c}^{3/2}}}^{2}][p^{2}-m_{_{P_{c \bar c}^{3/2}}}^{2}]}
 \Bigg[g_{\mu\nu}
-\frac{1}{3}\gamma_{\mu}\gamma_{\nu}-\frac{2\,p_{\mu}p_{\nu}}
{3\,m^{2}_{P_{c \bar c}^{3/2}}}+\frac{p_{\mu}\gamma_{\nu}-p_{\nu}\gamma_{\mu}}{3\,m_{P_{c \bar c}^{3/2}}}\Bigg]\nonumber\\
&\times \Bigg\{F_{1}(q^2)g_{\mu\nu}\eslash -
\frac{1}{2m_{P_{c \bar c}^{3/2}}}
\Big[F_{2}(q^2)g_{\mu\nu}+F_{4}(q^2) \frac{q_{\mu}q_{\nu}}{(2m_{P_{c \bar c}^{3/2}})^2}\Big]\eslash\qslash+\frac{F_{3}(q^2)}{(2m_{P_{c \bar c}^{3/2}})^2}
 q_{\mu}q_{\nu}\eslash\Bigg\}.
\end{align}

The above correlation function contains numerous Lorentz structures, not all of which are independent and this correlation function also includes spin-1/2 contributions. For our calculations to be more reliable, we need to get rid of these two problems. We can do this by choosing a specific  ordering for gamma matrices such as $\gamma_{\mu}\pslash\eslash\qslash\gamma_{\nu}$ 
 and abolish terms  with $\gamma_\mu$ at the beginning, $\gamma_\nu$ at the end and those proportional to $p_\mu$ and  $p_\nu$~\cite{Belyaev:1982cd}. 
After these algebraic manipulations, both spin-1/2 contributions are eliminated and all Lorentz structures become independent, and the final form of the hadronic side of the correlation function becomes the following form:
\begin{align}\label{final phenpart}
\Pi^{Had}_{\mu\nu}(p,q)&=\frac{\lambda_{_{{P_{c \bar c}^{3/2}}}}^{2}}{[(p+q)^{2}-m_{_{{P_{c \bar c}^{3/2}}}}^{2}][p^{2}-m_{_{{P_{c \bar c}^{3/2}}}}^{2}]} 
\Bigg[  g_{\mu\nu}\pslash\eslash\qslash \,F_{1}(q^2) 
-m_{{P_{c \bar c}^{3/2}}}g_{\mu\nu}\eslash\qslash\,F_{2}(q^2)
-
\frac{F_{3}(q^2)}{4m_{{P_{c \bar c}^{3/2}}}}q_{\mu}q_{\nu}\eslash\qslash
\nonumber\\
&
-
\frac{F_{4}(q^2)}{4m_{{P_{c \bar c}^{3/2}}}^3}(\varepsilon.p)q_{\mu}q_{\nu}\pslash\qslash 
+
\cdots 
\Bigg].
\end{align}

The magnetic dipole form factor, $G_{M}(q^2)$, of spin-3/2 pentaquarks can be written in terms of $F_{i}(q^2)$ form factors  as follows:~\cite{Weber:1978dh,Nozawa:1990gt,Pascalutsa:2006up,Ramalho:2009vc}:
\begin{align}
G_{M}(q^2) &= [ F_1(q^2) + F_2(q^2)] ( 1+ \frac{4}{5}
\tau ) -\frac{2}{5} [ F_3(q^2)  
+ 
F_4(q^2)] \tau ( 1 + \tau ), 
\end{align}
  where $\tau
= -\frac{q^2}{4m^2_{{P_{c \bar c}^{3/2}}}}$. At the static limit, i.e. $q^2=0$, the magnetic dipole moment is achieved in connection with the form factors as:
\begin{eqnarray}\label{mqo1}
G_{M}(q^2=0)&=&F_{1}(q^2=0)+F_{2}(q^2=0).
\end{eqnarray}
The  magnetic dipole moment of the spin-3/2 pentaquark state, ($\mu_{{P_{c \bar c}^{3/2}}}$), is defined  in the following way:
 \begin{eqnarray}\label{mqo2}
\mu_{{P_{c \bar c}^{3/2}}}&=&\frac{e}{2m_{{P_{c \bar c}^{3/2}}}}\big[F_{1}(q^2=0)+F_{2}(q^2=0)\big].
\end{eqnarray}

We have obtained the analytical expressions for both $P_{c\bar c}^{1/2}$ and $P_{c \bar c}^{3/2}$ pentaquarks. The next step in the calculations will be to perform numerical calculations of the analytical expressions obtained for the pentaquarks under investigation.

\end{widetext}

\section{Numerical Results}\label{numerical}

The  QCD light-cone sum rules for electromagnetic properties of pentaquarks without strange, with strange, and with double strange contain many input parameters that we need their numerical values.  Their numerical values are given as: $m_u=m_d=0$, $m_s =93.4^{+8.6}_{-3.4}\,\mbox{MeV}$, $m_c = 1.27 \pm 0.02\,\mbox{GeV}$~\cite{Workman:2022ynf},    $\langle \bar ss\rangle = 0.8\, \langle \bar uu\rangle$ $\,\mbox{GeV}^3$ with $\langle \bar uu\rangle =  \langle \bar dd\rangle=(-0.24 \pm 0.01)^3\,\mbox{GeV}^3$ \cite{Ioffe:2005ym}, $m_0^{2} = 0.8 \pm 0.1 \,\mbox{GeV}^2$ \cite{Ioffe:2005ym},  
$\langle g_s^2G^2\rangle = 0.88~ \mbox{GeV}^4$~\cite{Matheus:2006xi}, $f_{3\gamma}=-0.0039~\mbox{GeV}^2$~\cite{Ball:2002ps} and $\chi=-2.85 \pm 0.5~\mbox{GeV}^{-2}$~\cite{Rohrwild:2007yt}. 
To get numerical values for the electromagnetic multipole moments, we need to define the numerical values of the mass and residue of the pentaquarks. The masses and residues of these pentaquarks are borrowed from Refs.~\cite{Wang:2022neq, Wang:2022gfb}.
Photon DAs and their input parameters required for further analysis are taken from Ref.~\cite{Ball:2002ps}.

Predictions for electromagnetic properties extracted from the sum rules depend also on the Borel and continuum subtraction parameters $M^2$ and $s_0$. There should be a working interval where the results obtained will not vary much according to these extra parameters. The choice of working intervals for these extra parameters has to fulfill standard restrictions imposed on the pole contribution (PC) and convergence of the operator product expansion (OPE). It is convenient to use the following equations to describe these restrictions:
\begin{align}
 \mbox{PC} &=\frac{\Delta (M^2,s_0)}{\Delta (M^2,\infty)},\\
 \nonumber\\
 \mbox{OPE Convergence} &=\frac{\Delta^{\mbox{DimN}} (M^2,s_0)}{\Delta (M^2,s_0)},
 \end{align}
 where $\Delta^{\mbox{DimN}} (M^2,s_0)=\Delta^{\mbox{Dim(8+9+10)}} (M^2,s_0)$.   
 In the standard analysis of QCD light cone sum rules, pole contribution is expected to be over the $50\%$ for traditional hadrons. However, in multiquark states, this contribution is around $\mbox{PC}\geq 20\%$, which is enough for a reliable analysis.  To be convinced of the convergence of the OPE, we expect that these contributions should be less than $5\%$ of total calculations. The working regions obtained for $M^2$ and $s_0$ as a result of these restrictions are given in Table \ref{parameter} together with PC and convergence of OPE values for each state.  It follows from these values that the determined working regions for $M^2$ and $s_0$ meet the constraints coming from the dominance of PC and convergence of the OPE.  
 In Figs. \ref{Msqfig1} and \ref{Msqfig2}, we also illustrate the dependence of the magnetic dipole moment of pentaquarks, on the Borel mass parameter, $M^2$ at various values of $s_0$. It follows from these figures, that the variation of the respective magnetic dipole moments concerning $M^2$ is observed to be quite stable. When the variation of $s_0$ is examined, it is observed that the variation of the results according to this parameter is high, however, this variation remains within the error limits of the method used.
 
\begin{widetext}
 
 \begin{table}[htp]
	\addtolength{\tabcolsep}{10pt}
	\caption{Working intervals of  $s_0$ and  $M^2$ as well as the PC  and OPE convergence for the magnetic dipole moments of pentaquarks without strange, with strange, and with double strange.}
	\label{parameter}
		\begin{center}
\begin{tabular}{l|ccccc}
                \hline\hline
                \\
State & $s_0$ (GeV$^2$)& 
$M^2$ (GeV$^2$) & ~~  PC ($\%$) ~~ & ~~  OPE  
 ($\%$) \\
 \\
                                        \hline\hline
                                        \\
$\bar D \Xi_c^{\prime}$  & $23.5-25.5$ & $4.5-6.5$ & $30-55$ &  $2.8$  
                        \\
                        \\
$\bar D \Lambda_c$ & $23.6-25.6$ & $4.5-6.5$ & $31-56$ &  $2.9$  \\
                       \\
$\bar D_s \Xi_c$  & $24.4-26.4$ & $4.6-6.6$ & $30-57$ & $ 2.8$   \\
\\
$\bar D_s \Lambda_c$  & $23.8-25.8$ & $4.6-6.6$ & $32-57$ &  $2.5$   \\
                        \\
                                        \hline\hline
                                        \\
$\bar D^* \Xi_c^{\prime}$ & $24.7-26.7$ & $4.6-6.6$ & $31-54$ &  $2.8$  \\
                        \\
$\bar D^* \Lambda_c$ & $24.9-26.9$ & $4.6-6.6$ & $31-58$ &  $2.9$   \\
                        \\
$\bar D_s^* \Xi_c$   & $25.5-27.5$ &$4.7-6.7$ & $30-56$ & $2.7$   \\
\\
$\bar D_s^* \Lambda_c$ & $24.0-26.0$ & $4.5-6.5$ & $31-57$ &  $2.7$   \\
                       \\
                                       \hline\hline
 \end{tabular}
\end{center}
\end{table}

\end{widetext}

Employing all the input parameters we give the final results for the magnetic and higher multipole moments of pentaquarks in Table \ref{sonuc}. The presented uncertainties in the results originated from the errors in the values of all the input parameters as well as those errors coming from the calculations of the working regions for the extra parameters $M^2$ and $s_0$.

 In Ref.~\cite{Wang:2022tib}, the authors have studied the magnetic dipole moments of the $\bar D \Xi^{\prime}_c$ and $\bar D^{*}\Xi^{\prime}_c$ states in the framework of the constituent quark model and they predicted magnetic dipole moments as $-0.277$ and $-0.184$ for the $\bar D \Xi^{\prime}_c$ and $\bar D^{*}\Xi^{\prime}_c$ states, respectively. 
 When the quark model results are compared with the values obtained in this study, it is seen that the results are not compatible with each other.
In Ref.~\cite{Ozdem:2022iqk}, we obtained magnetic dipole moments of the hidden-charm pentaquarks with quantum number $J^P =1/2^-$ within the QCD light-cone sum rules. 
In the analyses, the interpolating currents $J^{10}(x)$, the light axialvector diquarks ([$q(x) C \gamma_\mu q(x)$]) combine with the heavy scalar diquarks ([$q(x) C \gamma_5 c(x)$]) to form a tetraquark with quantum number $J^P =1^+$, then this tetraquark couples with $\bar c$-quark ($[ \gamma_5 \gamma^\mu C \bar c(x)]$) to form a hidden-charm pentaquarks with quantum number $J^P =1/2^-$, while in the interpolating currents $J^{11}(x)$, the light axialvector diquarks ([$q(x) C \gamma_\mu q(x)$]) combine with the heavy axialvector diquarks ([$q(x) C \gamma^\mu c(x)$]) to form a tetraquark with quantum number $J^P =0^+$, then this tetraquark couples with $\bar c$-quark ($ [C \bar c(x)]$) to form a hidden-charm pentaquarks with quantum number $J^P =1/2^-$.  
As a result, in the calculations the axialvector-diquark-axialvector-diquark-antiquark ($P_{c \bar c}^{11}$) and axialvector-diquark-scalar-diquark-antiquark ($P_{c \bar c}^{10}$) forms of interpolating currents have been used to obtain the magnetic dipole moments of the hidden-charm pentaquarks without strange, with strange and with double strange.
The results obtained are given as follows: $\mu_{P_{c \bar cuds}^{11}}= 0.43^{+0.14}_{-0.16}~\mu_N$, $\mu_{P_{c \bar cuud}^{11}}= 0.60^{+0.25}_{-0.22}~\mu_N$, $\mu_{P_{c \bar cudd}^{11}}= 0.62^{+0.26}_{-0.22}~\mu_N$, $\mu_{P_{c \bar cdss}^{11}}= 0.70^{+0.25}_{-0.21}~\mu_N$,  
$\mu_{P_{c \bar cuds}^{10}}= 1.39^{+0.47}_{-0.44}~\mu_N$, $\mu_{P_{c \bar cuud}^{10}}= 1.42^{+0.57}_{-0.50}~\mu_N$, $\mu_{P_{c \bar cudd}^{10}}= 1.51^{+0.54}_{-0.49}~\mu_N$, $\mu_{P_{c \bar cdss}^{10}}= -1.92^{+0.65}_{-0.58}~\mu_N$.
It is seen that the results obtained for $\bar D^{0}\Lambda_c^+$, $\bar D^{-}\Lambda_c^+$, $\bar D_s^{-}\Xi_c^0$ and $\bar D_s^{-}\Lambda_c^+$ states are consistent with the axialvector-diquark-axialvector-diquark-antiquark configurations within errors. However, it can easily be seen that there is a large discrepancy with the results obtained using the axialvector-diquark-scalar-diquark-antiquark configuration. Using different models and configurations leads to different predictions. Therefore,  more theoretical investigation and experimental measurements are required to define the properties of these hidden-charm pentaquarks.
 Comparing the magnetic dipole moment results obtained using different theoretical models with the values obtained in this study may give an idea about the consistency of our estimations.
 
 \begin{widetext}

  \begin{table}[htp]
	\addtolength{\tabcolsep}{10pt}
	\caption{The magnetic dipole moments of hidden-charm pentaquarks without strange, with strange, and with double strange obtained by performing the QCD light-cone sum rules. For completeness, we have also presented higher multipole moments, electric quadrupole ($\mathcal{Q}$) and magnetic octupole ($\mathcal{O}$), of the $\bar D^{*}\Xi^{\prime}_c$, $\bar D^{*}\Lambda_c$,  $\bar D_s^{*}\Lambda_c$ and  $\bar D_s^{*}\Xi_c$ pentaquarks.}
	\label{sonuc}
		\begin{center}
\begin{tabular}{l|ccccc}
                \hline\hline
               \\
 Parameters& $\bar D \Xi_c^{\prime}$& 
$\bar D^0 \Lambda_c^+$ &$\bar D^- \Lambda_c^+$  & $\bar D_s^- \Xi_c^0$ & $\bar D_s^- \Lambda_c$ \\
    \\
                                        \hline\hline
                                      \\

 $\mu (\mu_N)$& $-0.10^{+0.03}_{-0.03}$ & $0.44^{+0.17}_{-0.14}$ & $0.45^{+0.17}_{-0.15}$ &  $0.50^{+0.18}_{-0.16}$  & $0.43^{+0.17}_{-0.14}$                        \\
                         \\
                                        \hline\hline
      \\  
       Parameters & $\bar D^* \Xi_c^{\prime}$&$\bar D^{*0} \Lambda_c^+$ &$\bar D^{*-} \Lambda_c^+$  & $\bar D_s^{*-} \Xi_c^0$ & $\bar D_s^{*-} \Lambda_c$ \\
      \\
 \hline \hline
 \\
$\mu (\mu_N)$ & $2.60^{+0.88}_{-0.77}$ & $2.24^{+0.77}_{-0.64}$ & $2.22^{+0.76}_{-0.64}$ &  $2.73^{+0.90}_{-0.81}$  & $1.87^{+0.70}_{-0.62}$  \\
                        \\
$\mathcal{Q}(\times 10^{-1}) (fm^2)$ & $1.60^{+0.40}_{-0.40}$ & $2.61^{+0.60}_{-0.60}$ & $2.50^{+0.60}_{-0.60}$ &  $3.11^{+0.70}_{-0.70}$  & $0.24^{+0.06}_{-0.06}$   \\
                        \\
$\mathcal{O}(\times 10^{-3})(fm^3)$   & $0.32^{+0.05}_{-0.05}$ & $0.11^{+0.04}_{-0.04}$ & $0.10^{+0.04}_{-0.04}$ &  $0.14^{+0.04}_{-0.04}$  & $0.09^{+0.03}_{-0.03}$   \\
                        \\
                                       \hline\hline
 \end{tabular}
\end{center}
\end{table}

 \end{widetext}
 
\section{summary and concluding remarks}\label{summary}
  
Since the discovery of the hidden-charm pentaquarks $P_c(4380)$ and $P_c(4450)$ by the LHCb Collaboration in 2015, the study and elucidation of pentaquarks, along with other pentaquark candidates discovered, has become an attractive subject in hadron physics. Inspired by this, we have studied the electromagnetic properties of $\bar D^{(*)}\Xi^{\prime}_c$, $\bar D^{(*)}\Lambda_c$,  $\bar D_s^{(*)}\Lambda_c$ and  $\bar D_s^{(*)}\Xi_c$  pentaquarks without strange, with strange and with double strange through QCD light-cone sum rules.  We have also evaluated electric quadrupole and magnetic octupole moments of the $\bar D^{*}\Xi^{\prime}_c$, $\bar D^{*}\Lambda_c$,  $\bar D_s^{*}\Lambda_c$ and  $\bar D_s^{*}\Xi_c$ pentaquarks. 
A comparison of our results on magnetic dipole moments of the hidden-charm pentaquarks with the other models existing in the literature is presented.
The magnetic dipole moments of the hidden-charm pentaquarks reveal valuable knowledge about the size and shape of the hadrons. Obtaining these parameters can be a prominent step in our understanding of hadron properties according to quark-gluon degrees of freedom. Moreover, the magnetic dipole moment is a key ingredient in the calculation of the $J/\psi$ photo-production procedure, which may provide an independent examination of the pentaquarks. It will also be crucial to identify the branching ratios of the different decay modes and decay channels of the hidden-charm pentaquarks.
With the accumulation of events, magnetic dipole moments of pentaquarks may be extracted from the comparison of theoretical and experimental cross-sections eventually in the near future. 
 If the inner structure of exotic states figures out, our comprehension of the construction of the subatomic world be crucially improved, and our comprehension of the non-perturbative behaviors of the strong interaction at the low energy region would also be crucially improved.

\begin{widetext} 

\begin{figure}[t]
\subfloat[]{\includegraphics[width=0.45\textwidth]{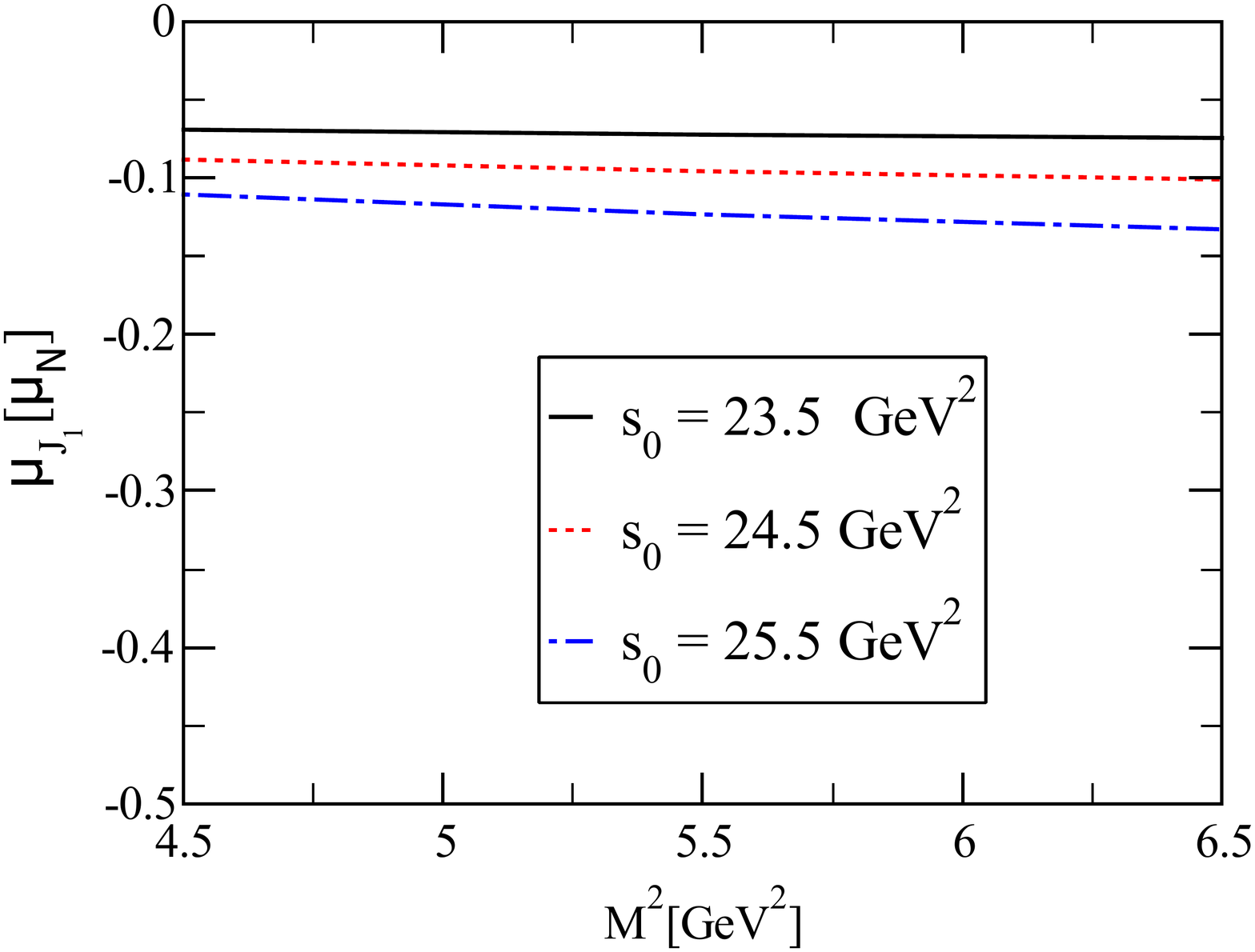}}~~~~
\subfloat[]{\includegraphics[width=0.45\textwidth]{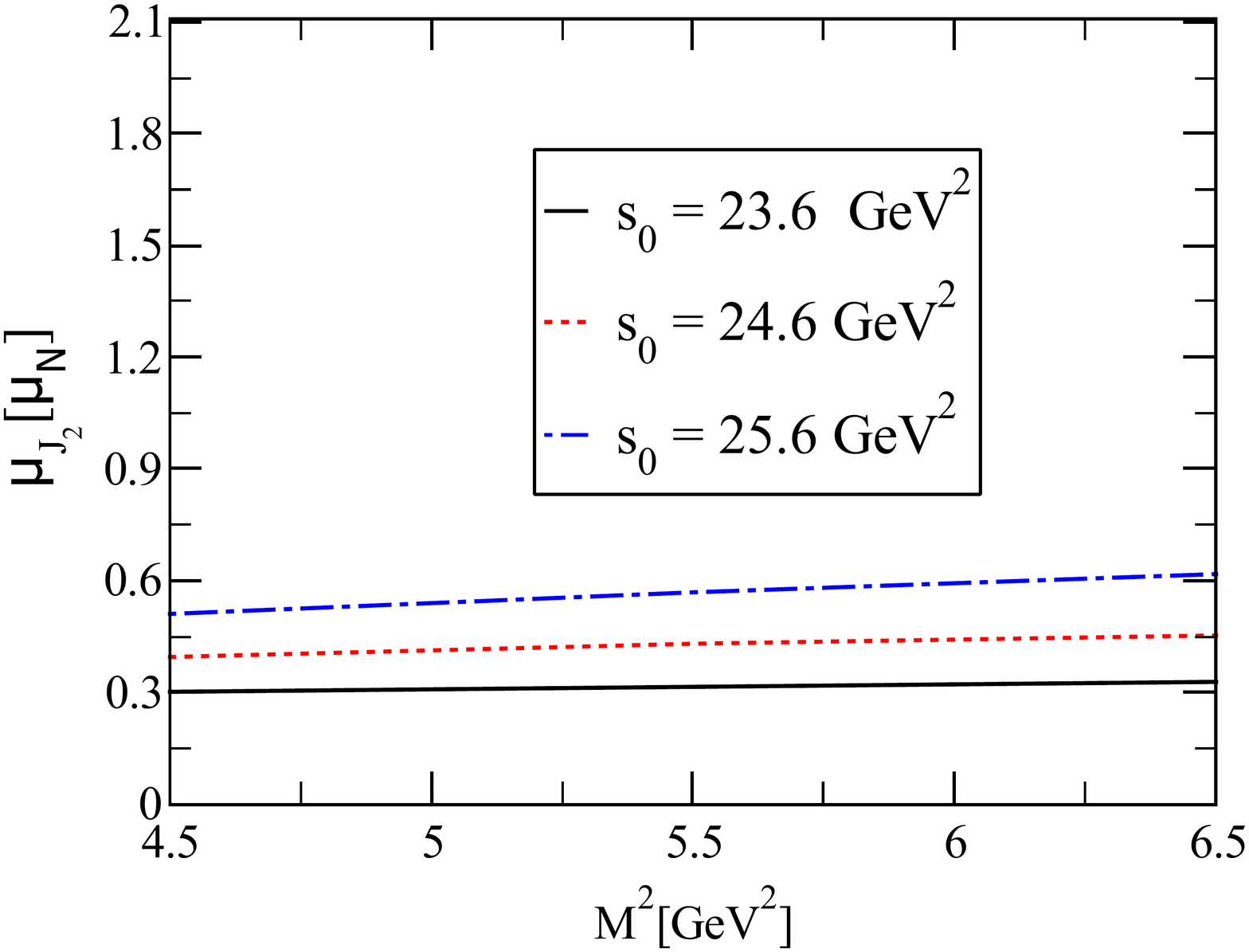}}\\
\subfloat[]{\includegraphics[width=0.45\textwidth]{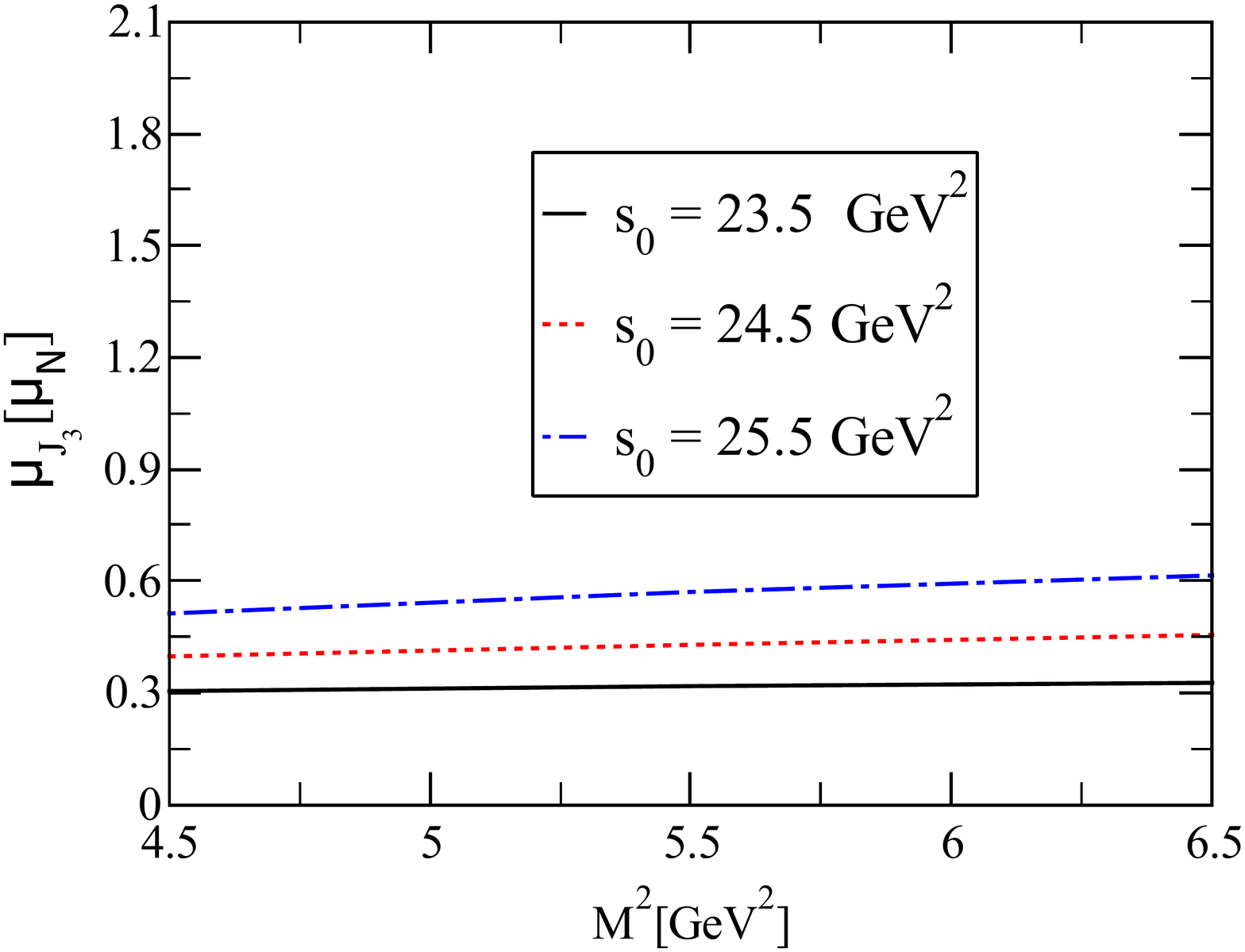}}~~~~
\subfloat[]{\includegraphics[width=0.45\textwidth]{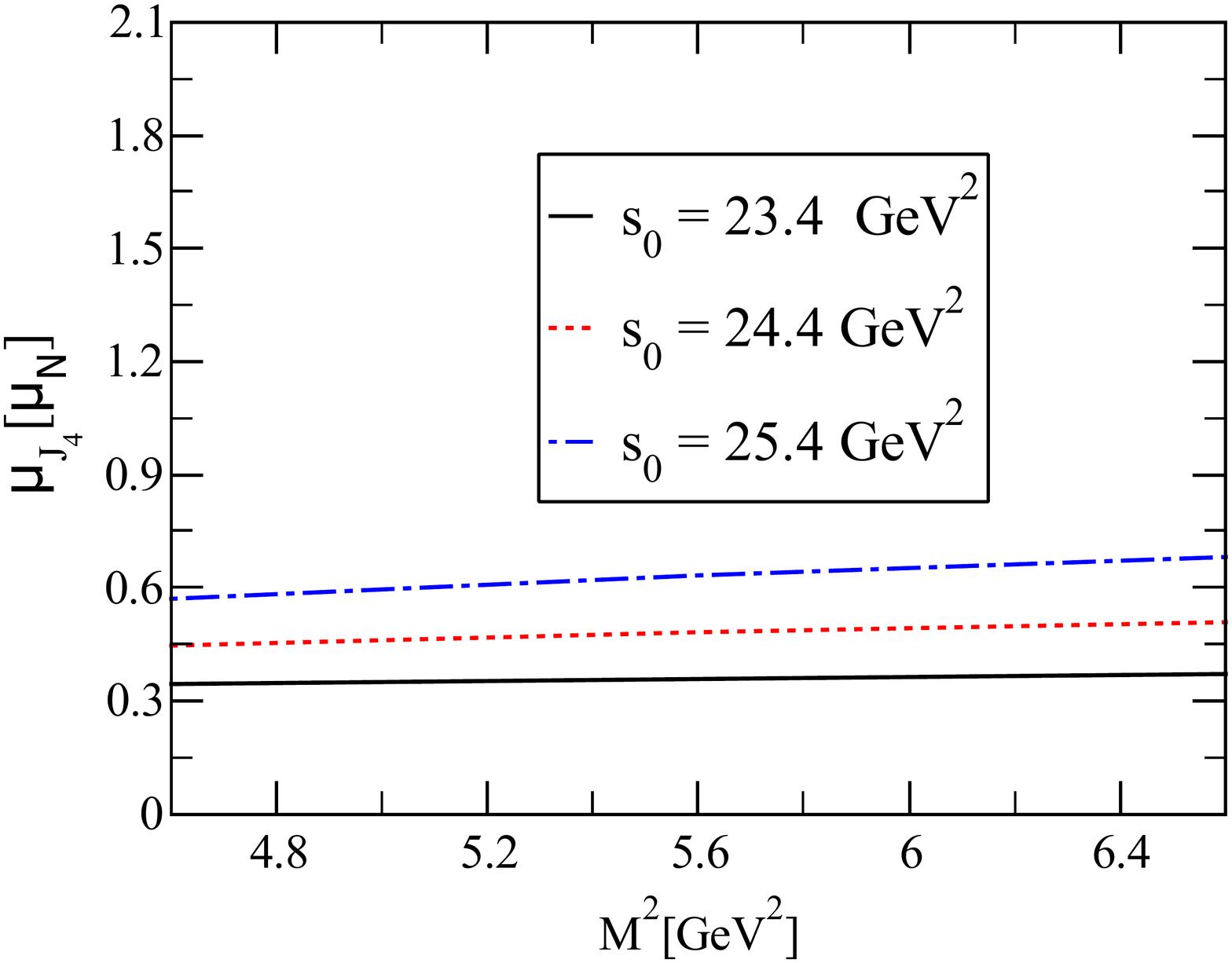}}\\
\subfloat[]{\includegraphics[width=0.45\textwidth]{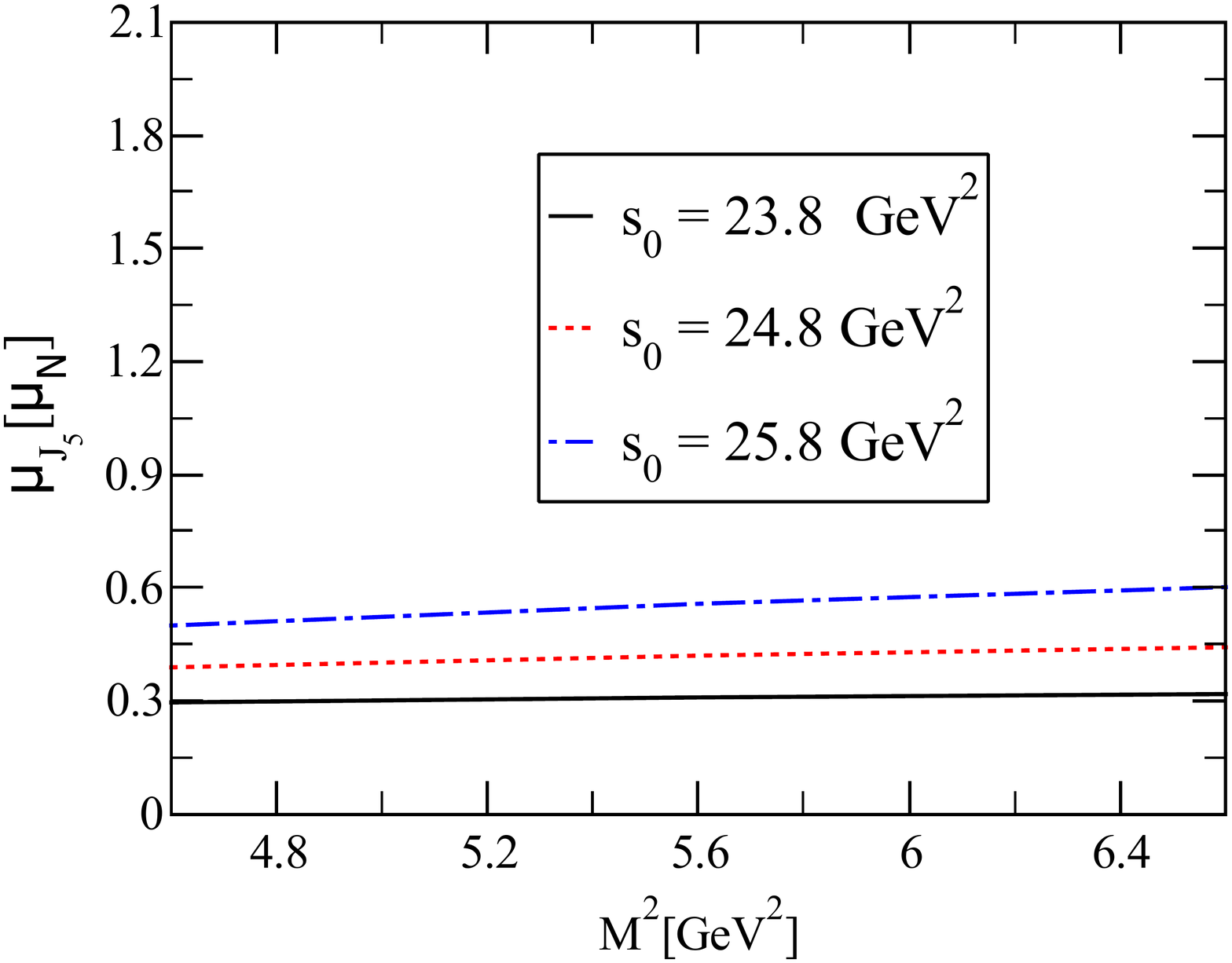}}
 \caption{The magnetic dipole moments of spin-$\frac{1}{2}$ pentaquarks versus $M^2$ at three different values of $s_0$; (a), (b), (c), (d) and (e) denote the $\bar D \Xi^{\prime}_c$, $\bar D^{0}\Lambda_c^+$, $\bar D^{-}\Lambda_c^+$, $\bar D_s^{-}\Xi_c^0$ and $\bar D_s^{-}\Lambda_c^+$    pentaquarks, respectively.}
 \label{Msqfig1}
  \end{figure}
  
%
 
 \begin{figure}[t]
\subfloat[]{\includegraphics[width=0.45\textwidth]{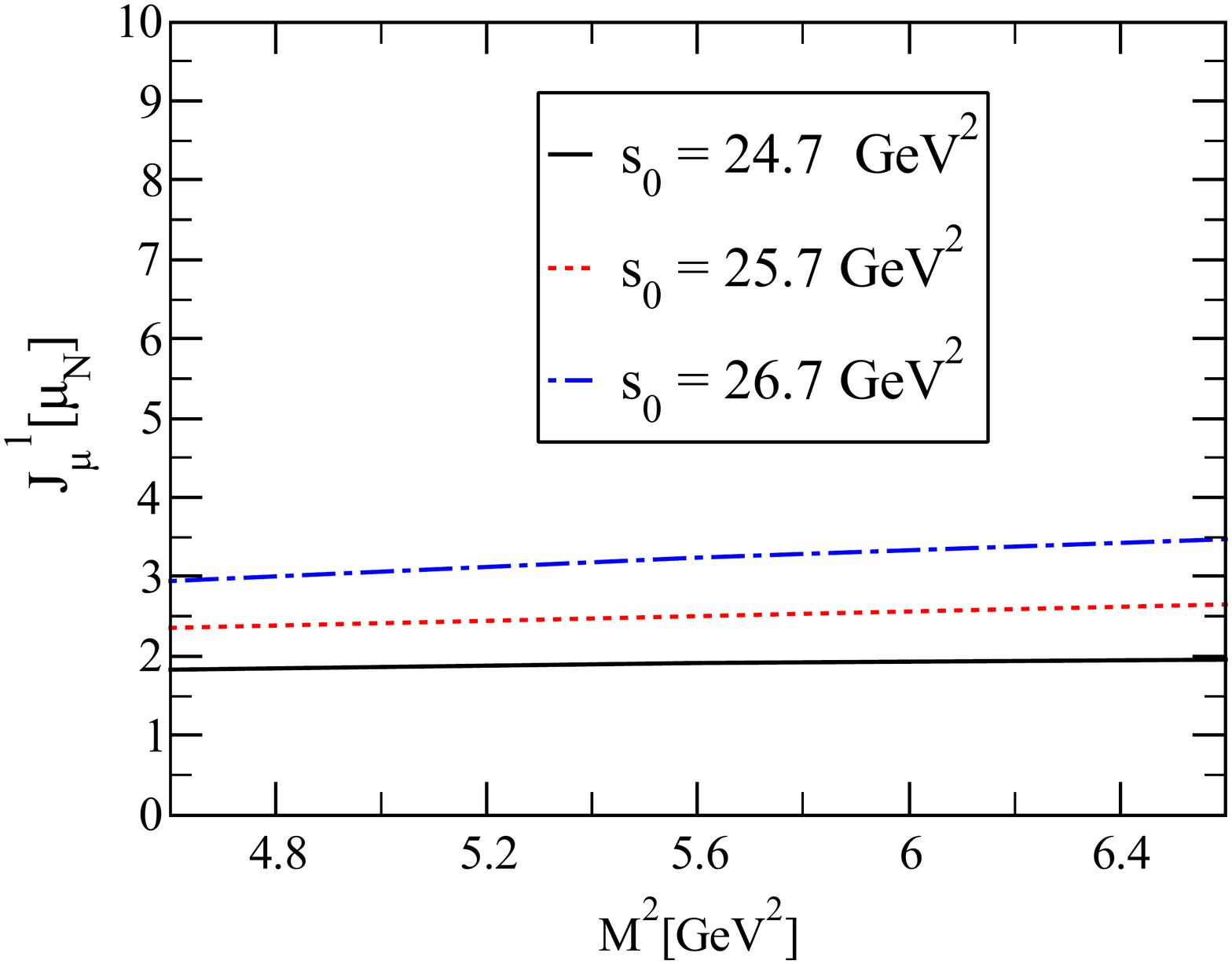}}~~~~
\subfloat[]{\includegraphics[width=0.45\textwidth]{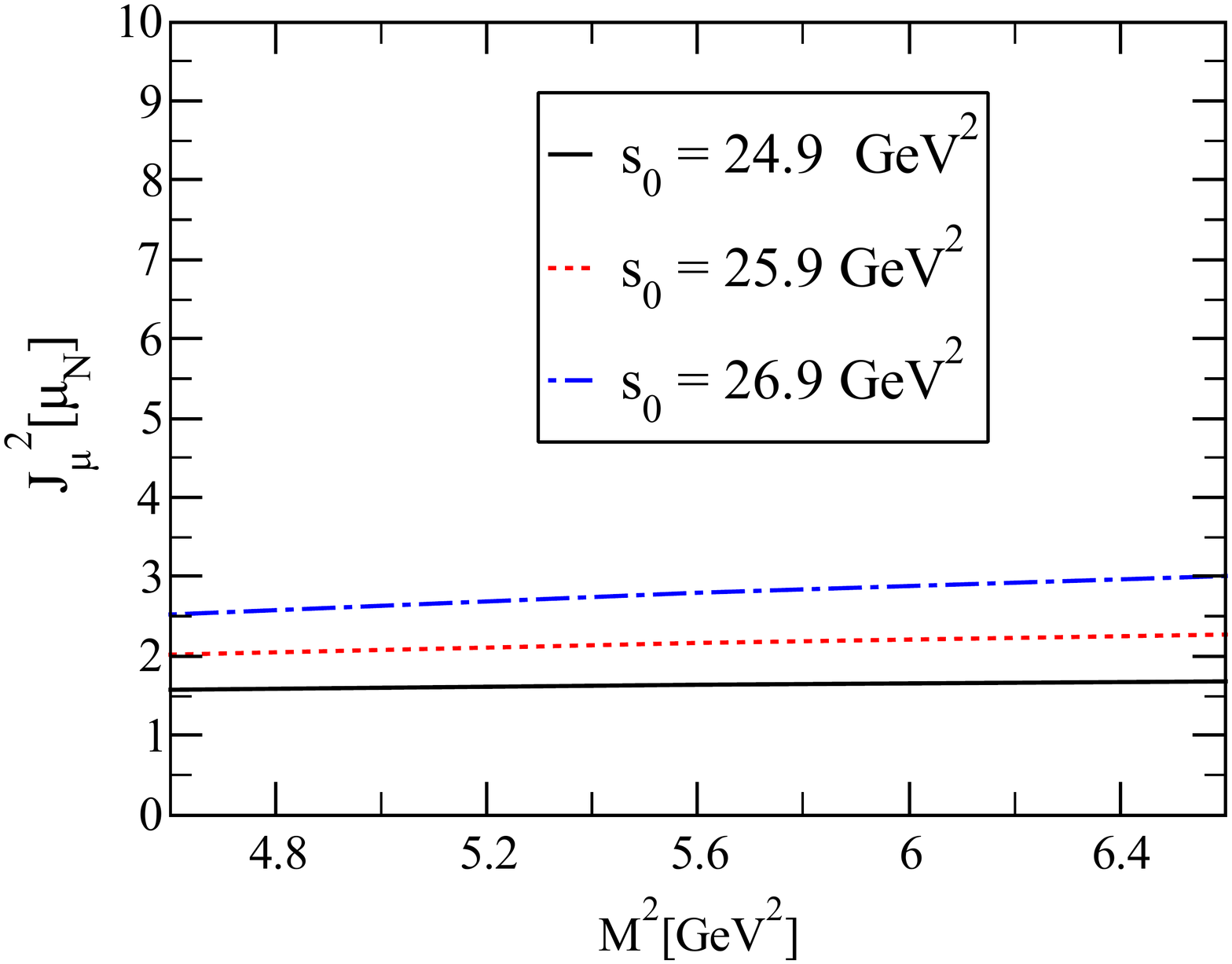}}\\
\subfloat[]{\includegraphics[width=0.45\textwidth]{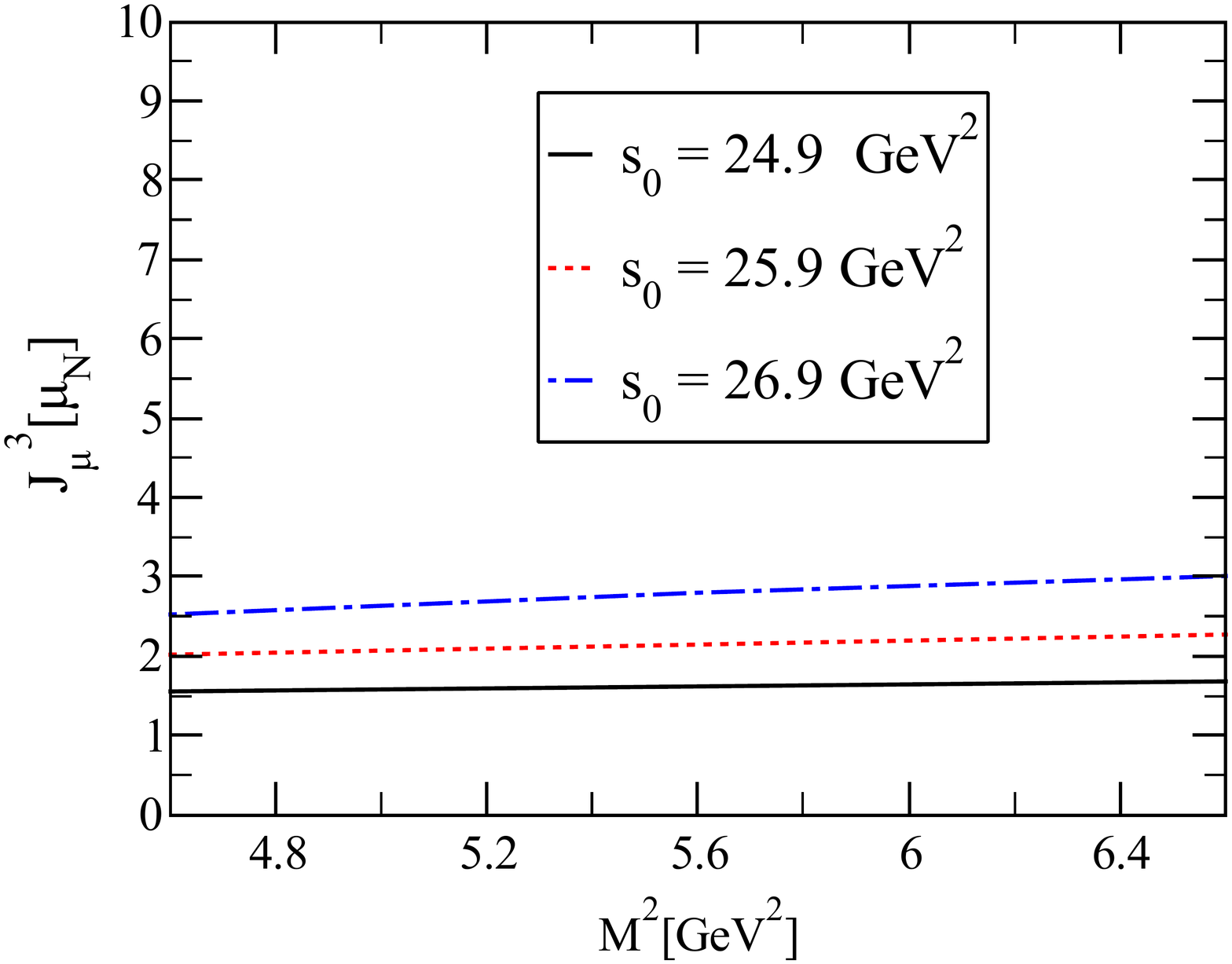}}~~~~
\subfloat[]{\includegraphics[width=0.45\textwidth]{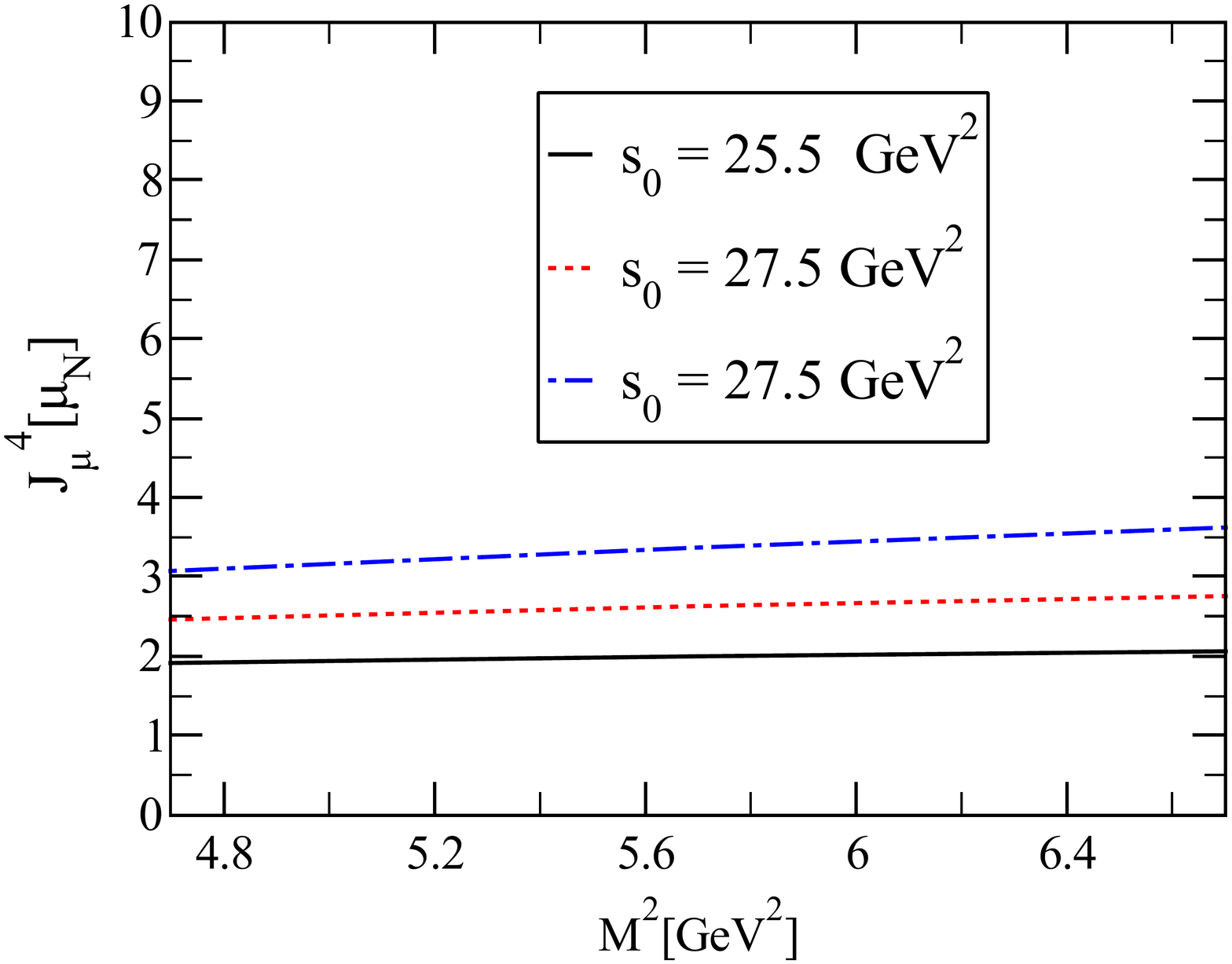}}\\
\subfloat[]{\includegraphics[width=0.45\textwidth]{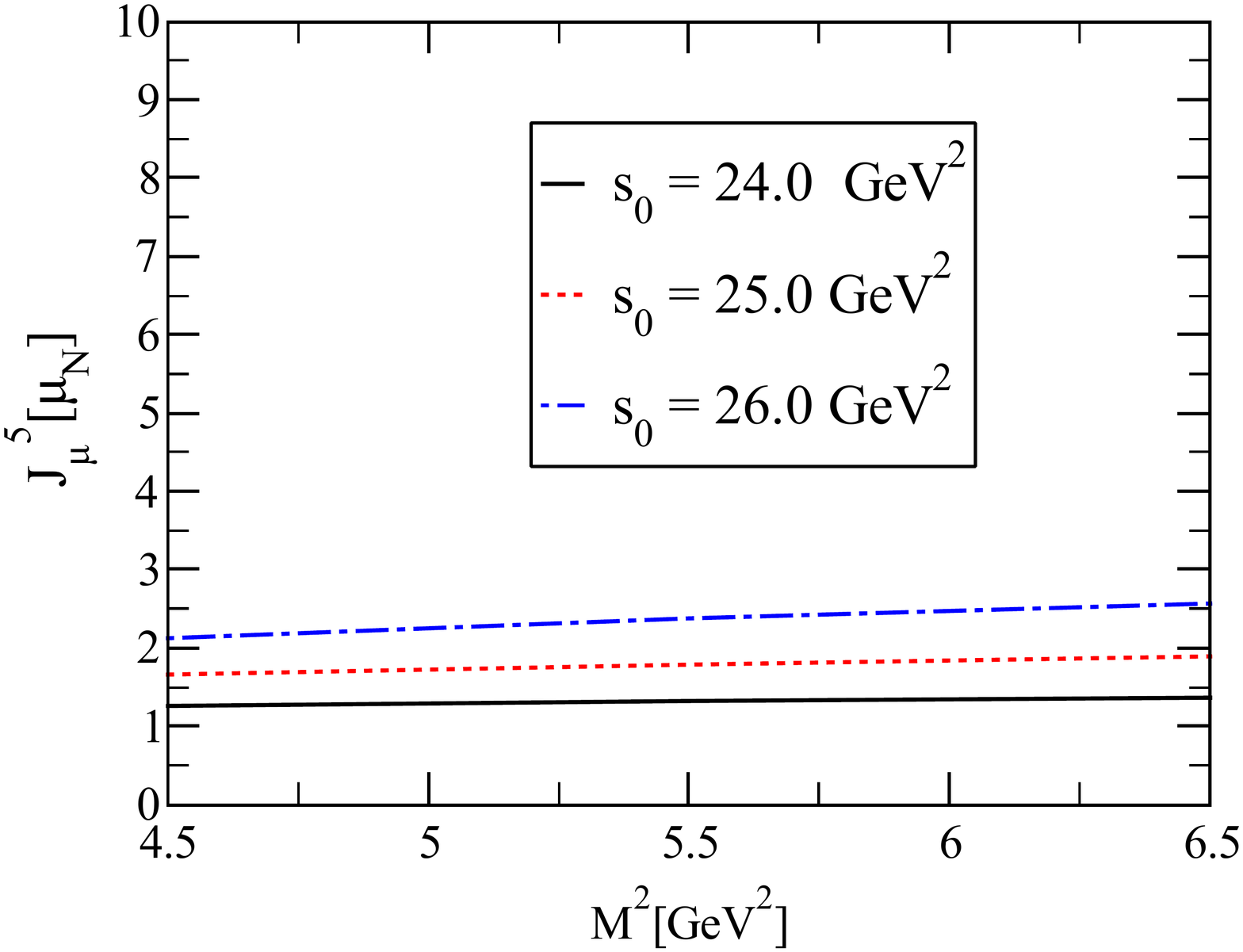}}
 \caption{The magnetic dipole moments of spin-$\frac{3}{2}$ pentaquarks versus $M^2$ at three different values of $s_0$; (a), (b), (c), (d) and (e) denote the $\bar D^{*}\Xi^{\prime}_c$, $\bar D^{*0}\Lambda_c^+$, $\bar D^{*-}\Lambda_c^+$, $\bar D_s^{*-}\Xi_c^0$ and $\bar D_s^{*-}\Lambda_c^+$    pentaquarks, respectively.}
 \label{Msqfig2}
  \end{figure}
 
   \end{widetext}


\begin{widetext}

\appendix
  \section*{Appendix: Explicit expression for \texorpdfstring{$\Delta_1^{QCD} (M^2,s_0)$}{}}\label{appenda}
 In this appendix, we present the explicit expressions of the function $\Delta_1^{QCD} (M^2,s_0)$ for the magnetic dipole moment of the $\bar D \Xi^{\prime}_c$ pentaquark state entering into the sum rule.
\begin{align}
 \Delta_1^{QCD}(M^2,s_0) &=\frac{m_c P_1 P_2 P_3  }{4076863488 \pi^3} (e_u + e_d)\Bigg\{
-3 (14 m_c m_s I[0, 1, 2, 0] + 19 I[0, 2, 3, 0]) I_ 2[\mathcal S] - 
 27 m_c m_s I_ 2[\mathcal {\tilde S}] I[0, 1, 2, 0] \nonumber\\
 &+ 
 96 m_c m_s I_ 6[h_ {\gamma}] I[0, 1, 2, 0] + 
 24 \Big ((I[0, 2, 1, 0] - 2 I[0, 2, 1, 1] + I[0, 2, 1, 2] - 
       2 I[0, 2, 2, 0] + 2 I[0, 2, 2, 1])
       \nonumber\\
 &\times \mathbb A[u_ 0] + (\mathbb A[u_ 0] + 
       I_ 5[\mathbb A] + \chi m_ 0^2 I_ 5[\varphi_ {\gamma}]) I[0, 2, 3, 
       0]\Big) - 
 8 \chi \big (3 m_c m_s I[0, 2, 3, 0] + 4 I[0, 3, 3, 0] 
 \nonumber\\
 &+ 
    3 I[0, 3, 4, 0]\big) I_ 5[\varphi_ {\gamma}] - 
 8 \chi \Big (3 m_c m_s \big (5 I[0, 2, 1, 0] - 6 I[0, 2, 1, 1] + 
        I[0, 2, 1, 2] - 6 I[0, 2, 2, 0] \nonumber\\
 &+ 2 I[0, 2, 2, 1] + 
        I[0, 2, 3, 0]\big) - 
     3 m_ 0^2 \big (I[0, 2, 1, 0] - 2 I[0, 2, 1, 1] + I[0, 2, 1, 2] - 
        2 I[0, 2, 2, 0] + 2 I[0, 2, 2, 1] \nonumber\\
 &+ I[0, 2, 3, 0]\big) - 
     7 I[0, 3, 1, 0] + 21 I[0, 3, 1, 1] - 21 I[0, 3, 1, 2] + 
     7 I[0, 3, 1, 3] + 17 I[0, 3, 2, 0] \nonumber\\
 &- 34 I[0, 3, 2, 1] + 
     17 I[0, 3, 2, 2] - 13 I[0, 3, 3, 0] + 13 I[0, 3, 3, 1] + 
     3 I[0, 3, 4, 0]\Big) \varphi_ {\gamma}[u_ 0]
 \Bigg\} \nonumber\\
 &-\frac { P_ 1 P_ 2^2 } {1019215872\pi^3}\Bigg\{48 m_c e_c \Big (I[0, 2, 1,
         0] - 2 I[0, 2, 1, 1] + I[0, 2, 1, 2] - 2 I[0, 2, 2, 0] + 
       2 I[0, 2, 2, 1]  \nonumber\\
 &- 
       10 \big (I[1, 1, 1, 0] - 2 I[1, 1, 1, 1] + I[1, 1, 1, 2] - 
           2 I[1, 1, 2, 0] + 2 I[1, 1, 2, 1] + I[1, 1, 3, 0]\big)\Big)
           \nonumber\\
 &
    + 96 m_c e_s \Big ( 
       3 I[0, 2, 1, 0] - 6 I[0, 2, 1, 1] + 3 I[0, 2, 1, 2] - 
        6 I[0, 2, 2, 0] + 6 I[0, 2, 2, 1] + 3 I[0, 2, 3, 0] 
        \nonumber\\
 &- 
        10 \big (I[1, 1, 1, 0] - 2 I[1, 1, 1, 1] + I[1, 1, 1, 2] - 
            2 I[1, 1, 2, 0] + 2 I[1, 1, 2, 1] + I[1, 1, 3, 0]\big)\Big)\nonumber\\
            & + (e_d + 
    e_u) m_c  \Bigg (-96 m_c m_s I_ 6[h_ {\gamma}] I[0, 1, 2, 0] + 
    3 I_ 2[\mathcal S] I[0, 2, 3, 0] + 
    12 \mathbb A[u_ 0] \Big (I[0, 2, 1, 0]\nonumber\\
            & - 2 I[0, 2, 1, 1] + 
       I[0, 2, 1, 2] - 2 I[0, 2, 2, 0] + 2 I[0, 2, 2, 1] + 
       I[0, 2, 3, 0]\Big) + 
    4 \chi \Big (24  m_c m_s (I[0, 2, 1, 0]\nonumber\\
            & - I[0, 2, 1, 1] - 
           I[0, 2, 2, 0]) + 
        3 m_ 0^2 (-2 I[0, 2, 1, 1] + I[0, 2, 1, 2] - 
           2 I[0, 2, 2, 0] + 2 I[0, 2, 2, 1] \nonumber\\
            &+ I[0, 2, 3, 0]) + 
        I[0, 3, 1, 0] - 3 I[0, 3, 1, 1] + 3 I[0, 3, 1, 2] - 
        I[0, 3, 1, 3] + I[0, 3, 2, 0] - 2 I[0, 3, 2, 1]\nonumber\\
            & + 
        I[0, 3, 2, 2] - 5 I[0, 3, 3, 0] + 5 I[0, 3, 3, 1] + 
        3 I[0, 3, 4, 0]\Big) \varphi_ {\gamma}[u_ 0] + 
    4 \Big (3 I_ 5[A] I[0, 2, 3, 
          0] + \chi I_ 5[\varphi_ {\gamma}] \nonumber\\
            & \times (3 m_ 0^2 I[0, 2, 3, 0] - 
           4 I[0, 3, 3, 0] + 3 I[0, 3, 4, 0]) + 
        3 \chi m_ 0^2 I[0, 2, 1, 0] \varphi_ {\gamma}[u_ 0]\Big)\Bigg)
      \Bigg\}\nonumber\\
 &     -\frac { P_ 1 P_ 2 } {5435817984 \pi^5}\Bigg\{48 m_c^2 m_0^2 \Bigg (8 \
e_s \Big (-3 I[0, 2, 2, 0] + 3 I[0, 2, 2, 1] + 3 I[0, 2, 3, 0] - 
          8 I[1, 1, 1, 0] 
          \nonumber\\
 &+ 2 I[1, 1, 1, 2] + 
          16 I[1, 1, 2, 0] - 6 I[1, 1, 2, 1] - 8 I[1, 1, 3, 0]\Big) + 
       e_c \Big (9 I[0, 2, 1, 0] - 54 I[0, 2, 1, 1] \nonumber\\
 &+ 
          45 I[0, 2, 1, 2] - 6 I[0, 2, 2, 0] + 42 I[0, 2, 2, 1] - 
          3 I[0, 2, 3, 0] - 
          2 (I[1, 1, 1, 0] + 6 I[1, 1, 1, 1] - 7 I[1, 1, 1, 2]\nonumber\\
 & - 
              2 I[1, 1, 2, 0] - 6 I[1, 1, 2, 1] + 
              I[1, 1, 3, 0])\Big) + 
       2 e_d \Big (I[0, 2, 1, 0] - 2 I[0, 2, 1, 1] + I[0, 2, 1, 2] - 
          8 I[0, 2, 2, 0] \nonumber\\
 &+ 8 I[0, 2, 2, 1] + 7 I[0, 2, 3, 0] - 
          4 (6 I[1, 1, 1, 0] - 11 I[1, 1, 1, 1] + 5 I[1, 1, 1, 2] - 
              12 I[1, 1, 2, 0] \nonumber\\
 &+ 11 I[1, 1, 2, 1] + 
              6 I[1, 1, 3, 0])\Big) + 
       2 e_u (I[0, 2, 1, 0] - 2 I[0, 2, 1, 1] + I[0, 2, 1, 2] - 
           8 I[0, 2, 2, 0] \nonumber\\
 &+ 8 I[0, 2, 2, 1] + 7 I[0, 2, 3, 0] - 
           4 (6 I[1, 1, 1, 0] - 11 I[1, 1, 1, 1] + 5 I[1, 1, 1, 2] - 
               12 I[1, 1, 2, 0] \nonumber\\
 &+ 11 I[1, 1, 2, 1] + 
               6 I[1, 1, 3, 0])\Big)\Bigg) - 
    192 e_c m_c^2 \Big (I[0, 3, 1, 0] + I[0, 3, 1, 1] - 9 I[0, 3, 1, 2] + 
       7 I[0, 3, 1, 3] \nonumber\\
 &- 2 I[0, 3, 2, 0] + 6 I[0, 3, 2, 2] + 
       I[0, 3, 3, 0] - I[0, 3, 3, 1] - I[1, 2, 1, 1] - 
       2 I[1, 2, 1, 2] + 3 I[1, 2, 1, 3]\nonumber\\
 & + 
       2 (I[1, 2, 2, 1] + I[1, 2, 2, 2]) - I[1, 2, 3, 1]\Big) + 
    576 e_s m_c^2 \Big (2 I[0, 3, 1, 0] - 4 I[0, 3, 1, 1] + 3 I[0, 3, 1, 2] \nonumber
  \end{align}
 
 \begin{align}
 &-
        I[0, 3, 1, 3] - 4 I[0, 3, 2, 0] + 6 I[0, 3, 2, 1] - 
       3 I[0, 3, 2, 2] + 2 I[0, 3, 3, 0] - 2 I[0, 3, 3, 1] + 
       8 I[1, 2, 1, 1] \nonumber\\
 &- 11 I[1, 2, 1, 2] + 3 I[1, 2, 1, 3] - 
       16 I[1, 2, 2, 1] + 11 I[1, 2, 2, 2] + 8 I[1, 2, 3, 1]\Big) + 
    144 e_d m_c^2 \Big (7 I[0, 3, 1, 0] \nonumber\\
 &- 15 I[0, 3, 1, 1] + 
       12 I[0, 3, 1, 2] - 4 I[0, 3, 1, 3] - 14 I[0, 3, 2, 0] + 
       22 I[0, 3, 2, 1] - 11 I[0, 3, 2, 2] \nonumber\\
  &+ 7 I[0, 3, 3, 0] - 
       7 I[0, 3, 3, 1] + 36 I[1, 2, 1, 1] - 69 I[1, 2, 1, 2] + 
       33 I[1, 2, 1, 3] - 72 I[1, 2, 2, 1]\nonumber\\
 & + 69 I[1, 2, 2, 2] + 
       36 I[1, 2, 3, 1]\Big) + 
    144 e_u m_c^2 \Big (7 I[0, 3, 1, 0] - 15 I[0, 3, 1, 1] + 
        12 I[0, 3, 1, 2] - 4 I[0, 3, 1, 3] \nonumber\\
 & - 14 I[0, 3, 2, 0] + 
        22 I[0, 3, 2, 1] - 11 I[0, 3, 2, 2] + 7 I[0, 3, 3, 0] - 
        7 I[0, 3, 3, 1] + 36 I[1, 2, 1, 1] \nonumber\\
 &- 69 I[1, 2, 1, 2] + 
        33 I[1, 2, 1, 3] - 72 I[1, 2, 2, 1] + 69 I[1, 2, 2, 2] + 
        36 I[1, 2, 3, 1]\Big)
        \nonumber\\
        &
        +4 m_c \Bigg (3 \chi m_s (-e_u I[0, 4, 4, 0] + (e_d + e_u) I[0, 4, 4, 
         1]) \varphi_ {\gamma}[u_ 0] - 
   16 f_ {3\gamma} \pi^2\Bigg (2 e_s m_c \Big (2 m_ 0^2 (I[0, 1, 1, 
             0] - I[0, 1, 1, 1] \nonumber\\
        &- I[0, 1, 2, 0]) + 11 I[0, 2, 1, 0] - 
         22 I[0, 2, 1, 1] + 11 I[0, 2, 1, 2] - 18 I[0, 2, 2, 0] + 
         18 I[0, 2, 2, 1] + 7 I[0, 2, 3, 0]\Big) 
         \nonumber\\
        &+ 
      e_d \Big (2 m_ 0^2 m_c (I[0, 1, 1, 0] - I[0, 1, 1, 1] - 
            I[0, 1, 2, 0]) + 
         2  m_s \big (I[0, 2, 1, 0] - 2 I[0, 2, 1, 1] + 
            I[0, 2, 1, 2] - 2 I[0, 2, 2, 0] 
            \nonumber\\
        &+ 2 I[0, 2, 2, 1] + 
            I[0, 2, 3, 0]\big) - 
         m_c\big (3 I[0, 2, 1, 0] - 6 I[0, 2, 1, 1] + 
             3 I[0, 2, 1, 2] - 10 I[0, 2, 2, 0] + 10 I[0, 2, 2, 1] 
             \nonumber\\
        &+ 
             7 I[0, 2, 3, 0]\big)\Big) + 
      e_u \Big (2 m_ 0^2 m_c (I[0, 1, 1, 0] - I[0, 1, 1, 1] - 
             I[0, 1, 2, 0]) + 
          2  m_s \big (I[0, 2, 1, 0] - 2 I[0, 2, 1, 1] + 
             I[0, 2, 1, 2] 
             \nonumber\\
        &- 2 I[0, 2, 2, 0] + 2 I[0, 2, 2, 1] + 
             I[0, 2, 3, 0]\big) - 
          m_c\big (3 I[0, 2, 1, 0] - 6 I[0, 2, 1, 1] + 
              3 I[0, 2, 1, 2] - 10 I[0, 2, 2, 0]  
              \nonumber\\
        &+ 10 I[0, 2, 2, 1]+ 
              7 I[0, 2, 3, 0]\big)\Big)\Bigg) \psi^a[u_ 0]\Bigg)
              +
              4 \chi m_c (e_d + 
    e_u) \Bigg (4 m_c \Big (-I[0, 4, 1, 2] + I[0, 4, 1, 3] + 
       I[0, 4, 2, 2]\Big) \nonumber\\
        &+ 
    m_s \Big (3 I[0, 4, 1, 0] - 8 I[0, 4, 1, 1] + 8 I[0, 4, 1, 2] - 
        4 I[0, 4, 1, 3] + I[0, 4, 1, 4] - 9 I[0, 4, 2, 0] + 
        19 I[0, 4, 2, 1] 
        \nonumber\\
        &- 15 I[0, 4, 2, 2] + 5 I[0, 4, 2, 3] + 
        9 I[0, 4, 3, 0] - 14 I[0, 4, 3, 1] + 7 I[0, 4, 3, 2] - 
        3 I[0, 4, 4, 0] + 
        3 I[0, 4, 4, 1]\Big)\Bigg) \varphi_ {\gamma}[u_ 0]\nonumber\\
        &
        +64 f_ {3\gamma} m_c \Bigg ((-5 e_s - 2 e_u) m_c I_ 2[\mathcal V] I[0,
      2, 2, 0] + \Big (2 (e_d + 2 e_s + 
         e_u) m_c (m_ 0^2 I[0, 1, 2, 0] - 
         2 I[0, 2, 2, 0]) \nonumber\\
        &+ (7 (e_d - 2 e_s + e_u) m_c + 
         2 (e_d + e_u) m_s) I[0, 2, 3, 0]\Big) I_5[\psi^a] \Bigg)
      \Bigg\}\nonumber\\
              &
              + \frac {P_ 1 P_ 3} {16307453952  \pi^5}\Bigg\{-96 m_0^2\Bigg(3  (3 e_d - e_c + 3 e_u) m_c^2 (I[0, 2, 1, 0] - 2 I[0, 2, 1, 1] + 
          I[0, 2, 1, 2] - 2 I[0, 2, 2, 0] 
          \nonumber\\
        &+ 2 I[0, 2, 2, 1] + 
          I[0, 2, 3, 0] - 2 I[1, 1, 1, 0] + 4 I[1, 1, 1, 1] - 
          2 I[1, 1, 1, 2] + 4 I[1, 1, 2, 0] - 4 I[1, 1, 2, 1] - 
          2 I[1, 1, 3, 0])
          \nonumber\\
        &
       + m_c m_s \Big (e_u (I[0, 2, 1, 0] - 2 I[0, 2, 1, 1] + 
              I[0, 2, 1, 2] - 2 I[0, 2, 2, 0] + 2 I[0, 2, 2, 1] + 
              2 I[1, 1, 1, 0] - 4 I[1, 1, 1, 1] 
              \nonumber\\
        &+ 2 I[1, 1, 1, 2] - 
              4 I[1, 1, 2, 0] + 4 I[1, 1, 2, 1] + 2 I[1, 1, 3, 0]) + 
           e_d (I[0, 2, 1, 0] - 2 I[0, 2, 1, 1] + I[0, 2, 1, 2] - 
              2 I[0, 2, 2, 0]
              \nonumber\\
        &+ 2 I[0, 2, 2, 1] + I[0, 2, 3, 0] + 
              2 I[1, 1, 1, 0] - 4 I[1, 1, 1, 1] + 2 I[1, 1, 1, 2] - 
              4 I[1, 1, 2, 0] + 4 I[1, 1, 2, 1] + 2 I[1, 1, 3, 0]) 
              \nonumber\\
        &+ 
           e_c (25 I[0, 2, 1, 0] - 50 I[0, 2, 1, 1] + 
               25 I[0, 2, 1, 2] - 50 I[0, 2, 2, 0] + 
               50 I[0, 2, 2, 1] + 25 I[0, 2, 3, 0] + 
               2 I[1, 1, 1, 0] \nonumber\\
        &- 4 I[1, 1, 1, 1] + 2 I[1, 1, 1, 2] - 
               4 I[1, 1, 2, 0] + 4 I[1, 1, 2, 1] + 
               2 I[1, 1, 3, 0])\Big)\Bigg) + 
    96 m_s (e_d + e_u + 3 e_c)  \Big (I[0, 3, 1, 0] \nonumber\\
        &- 
       3 I[0, 3, 1, 1] + 3 I[0, 3, 1, 2] - I[0, 3, 1, 3] - 
       41 I[0, 3, 2, 0] + 82 I[0, 3, 2, 1] - 41 I[0, 3, 2, 2] + 
       79 I[0, 3, 3, 0] \nonumber\\
        &- 79 I[0, 3, 3, 1] - 
       3 (13 I[0, 3, 4, 0] - 5 I[1, 2, 1, 1] + I[1, 2, 1, 2] + 
           I[1, 2, 1, 3] + 94 I[1, 2, 2, 1] - 43 I[1, 2, 2, 2]\nonumber\\
        & - 
           89 I[1, 2, 3, 1])\Big) - 
    576 m_c (5 e_d + 7 e_u + 3 e_c) \Big (I[0, 3, 1, 0] - 
        3 I[0, 3, 1, 1] + 3 I[0, 3, 1, 2] - I[0, 3, 1, 3] - 
        41 I[0, 3, 2, 0] \nonumber\\
        &+ 82 I[0, 3, 2, 1] - 41 I[0, 3, 2, 2] + 
        79 I[0, 3, 3, 0] - 79 I[0, 3, 3, 1] - 
        3 (13 I[0, 3, 4, 0] - 5 I[1, 2, 1, 1] + I[1, 2, 1, 2] 
        \nonumber\\
        &+ 
            I[1, 2, 1, 3] + 94 I[1, 2, 2, 1] - 43 I[1, 2, 2, 2] - 
            89 I[1, 2, 3, 1])\Big)
            -3 f_{3\gamma} m_c \pi^2  \Bigg (256 e_d m_c  (m_ 0^2 I[0, 1, 2, 
         0] - 2 I[0, 2, 2, 
         0])\nonumber\\
        & \times I_6[\psi_{\gamma}^{\nu}] + (184 e_u m_c I[0, 2, 2, 
         0] - 5 e_d m_s I[0, 2, 3, 0] - 
       5  e_u m_s I[0, 2, 3, 0]) I_ 2[\mathcal V] + 
    64 (e_d + 
        e_u) \Big (- 
          m_s (I[0, 2, 1, 0] 
          \nonumber
       \end{align}
\begin{align}
  &- 2 I[0, 2, 1, 1] + I[0, 2, 1, 2] - 
           2 I[0, 2, 2, 0] + 2 I[0, 2, 2, 1] + I[0, 2, 3, 0]) + 
        m_c (5 I[0, 2, 1, 0] \nonumber\\
        &- 10 I[0, 2, 1, 1] + 
            5 I[0, 2, 1, 2] - 6 I[0, 2, 2, 0] + 6 I[0, 2, 2, 1] + 
            I[0, 2, 3, 0])\Big) \psi^a[u_ 0]\Bigg) + 
 12 e_s m_c^2 \Bigg (-(4 I[0, 3, 2, 0] \nonumber\\
        &+ 
        I[0, 3, 3, 0]) I_ 2[\mathcal S] - 
    4 \Bigg (2 I_ 4[\mathcal {\tilde S}] I[0, 3, 2, 0] + 
       2 I_ 5[\mathcal A] I[0, 3, 2, 0] - 
       2 I_ 6[h_{\gamma}] I[0, 3, 2, 0] + 
       I_ 5[\mathcal A] I[0, 3, 3, 0] \nonumber\\
        &+ 
       I_ 4[\mathcal S] (8 I[0, 3, 2, 0] + I[0, 3, 3, 0])\Big) + 
    4  \Big (I[0, 3, 1, 0] - 5 I[0, 3, 1, 1] + 7 I[0, 3, 1, 2] - 
       3 I[0, 3, 1, 3] - 2 I[0, 3, 2, 0] \nonumber\\
        &+ 6 I[0, 3, 2, 1] - 
       4 I[0, 3, 2, 2] + I[0, 3, 3, 0] - I[0, 3, 3, 1]\Big) \mathbb A[
       u_ 0]\Bigg) - 
 8 \chi e_s m_c^2 \big(6 I[0, 4, 1, 1] - 21 I[0, 4, 1, 2] \nonumber\\
        &+ 
     22 I[0, 4, 1, 3] - 7 I[0, 4, 1, 4] - 12 I[0, 4, 2, 1] + 
     24 I[0, 4, 2, 2] - 10 I[0, 4, 2, 3] + 6 I[0, 4, 3, 1] - 
     3 I[0, 4, 3, 2]\Big) \varphi_{\gamma}[u_ 0]
      \Bigg\}\nonumber\\
   &-\frac{m_c P_2 P_3}{18874368 \pi^3}\Bigg\{ 
   \big(4 (5 e_d - e_u) m_c m_s I[0, 3, 3, 0] - 
    21 (e_d - 2 e_s + e_u) I[0, 4, 4, 0]\big) I_ 4[\mathcal S] + 
 4 (e_d + e_u) \Bigg(-3 I[0, 4, 2, 0] 
 \nonumber\\
        &+ 9 I[0, 4, 2, 1] - 
    9 I[0, 4, 2, 2] + 3 I[0, 4, 2, 3] + 9 I[0, 4, 3, 0] - 
    18 I[0, 4, 3, 1] + 9 I[0, 4, 3, 2] - 9 I[0, 4, 4, 0] + 
    9 I[0, 4, 4, 1] 
    \nonumber\\
        &
        + 3 I[0, 4, 5, 0] + 
    28 m_c m_s \Big (I[0, 3, 2, 0] - 2 I[0, 3, 2, 1] + 
       I[0, 3, 2, 2] - 2 I[0, 3, 3, 0] + 2 I[0, 3, 3, 1] + 
       I[0, 3, 4, 0] 
       \nonumber\\
        &+ 6 I[1, 2, 2, 1] - 3 I[1, 2, 2, 2] - 
       6 I[1, 2, 3, 1]\Big) - 
    12 \Big (3 I[1, 3, 2, 1] - 3 I[1, 3, 2, 2] + I[1, 3, 2, 3] + 
        -6 I[1, 3, 3, 1]
        \nonumber\\
        &+ 3I[1, 3, 3, 2] + 
            3I[1, 3, 4, 1]\Big)\Bigg) - 
 16 e_c \Bigg(m_0^2 \Big (5 m_c m_s \big(3 I[0, 2, 1, 0] - 
          6 I[0, 2, 1, 1] + 3 I[0, 2, 1, 2] - 6 I[0, 2, 2, 0] 
          \nonumber\\
        &+ 
          6 I[0, 2, 2, 1] + 3 I[0, 2, 3, 0] + 
          2I[1, 1, 1, 0] - 4 I[1, 1, 1, 1] + 2I[1, 1, 1, 2] - 
              4 I[1, 1, 2, 0] + 4 I[1, 1, 2, 1] + 
             2 I[1, 1, 3, 0]\big) 
             \nonumber\\
        &- 
       2  \big (I[0, 3, 2, 0] - 2 I[0, 3, 2, 1] + I[0, 3, 2, 2] - 
           2 I[0, 3, 3, 0] + 2 I[0, 3, 3, 1] + I[0, 3, 4, 0] + 
           6 I[1, 2, 2, 1] - 3 I[1, 2, 2, 2] 
           \nonumber\\
        &- 
           6 I[1, 2, 3, 1]\big)\Big) + 
    4 m_c m_s \Big(I[0, 3, 1, 0] - 5 I[0, 3, 1, 1] + 
       7 I[0, 3, 1, 2] - 3 I[0, 3, 1, 3] - 2 I[0, 3, 2, 0] + 
       8 I[0, 3, 2, 1] 
       \nonumber\\
        &- 6 I[0, 3, 2, 2] + I[0, 3, 3, 0] - 
       3 \big (I[0, 3, 3, 1] + I[1, 2, 1, 1] - 2 I[1, 2, 1, 2] + 
           I[1, 2, 1, 3] - 2 I[1, 2, 2, 1] + 2 I[1, 2, 2, 2] 
           \nonumber\\
        &+ 
           I[1, 2, 3, 1]\big)\Big) - I[0, 4, 2, 0] - 
    3 I[0, 4, 2, 1] + 3 I[0, 4, 2, 2] - I[0, 4, 2, 3] - 
    2 I[0, 4, 3, 0] + 4 I[0, 4, 3, 1] - 2 I[0, 4, 3, 2] \nonumber\\
        &+ 
    I[0, 4, 4, 0] - I[0, 4, 4, 1] + 4 I[1, 3, 2, 1] - 
    8 I[1, 3, 2, 2] + 4 I[1, 3, 2, 3] - 8 I[1, 3, 3, 1] + 
    8 I[1, 3, 3, 2] + 4 I[1, 3, 4, 1]\Bigg)   \Bigg\}\nonumber\\
    &
    - \frac {m_c P_ 2^2} {18874368 \pi^3}\Bigg\{\Big (8 e_d m_c m_s I[0, 
         3, 3, 0] + 56 e_u m_c m_s I[0, 3, 3, 0] - 
       3 e_d I[0, 4, 4, 0] - 
       3 e_u I[0, 4, 4, 0]\Big) I_ 4[\mathcal S] \nonumber\\
        &- 
    8 e_c \Bigg (8 m_c m_s \Big (I[0, 3, 1, 0] - 2 I[0, 3, 1, 1] + 
          I[0, 3, 1, 2] - 2 I[0, 3, 2, 0] + 2 I[0, 3, 2, 1] + 
          I[0, 3, 3, 0]\Big) + 3 I[0, 4, 1, 0] 
          \nonumber\\
        &- 15 I[0, 4, 1, 1] + 
       27 I[0, 4, 1, 2] - 21 I[0, 4, 1, 3] + 6 I[0, 4, 1, 4] - 
       10 I[0, 4, 2, 0] + 39 I[0, 4, 2, 1] - 48 I[0, 4, 2, 2]
       \nonumber\\
        &+ 
       19 I[0, 4, 2, 3] + 11 I[0, 4, 3, 0] - 31 I[0, 4, 3, 1] + 
       20 I[0, 4, 3, 2] - 4 I[0, 4, 4, 0] + 7 I[0, 4, 4, 1] + 
       2 m_ 0^2 \Big (6 I[0, 3, 1, 0] \nonumber\\
        &- 18 I[0, 3, 1, 1] + 
          18 I[0, 3, 1, 2] - 6 I[0, 3, 1, 3] - 19 I[0, 3, 2, 0] + 
          38 I[0, 3, 2, 1] - 19 I[0, 3, 2, 2] + 20 I[0, 3, 3, 0] 
          \nonumber\\
        &- 
          20 I[0, 3, 3, 1] - 7 I[0, 3, 4, 0] - 6 I[1, 2, 2, 1] + 
          3 I[1, 2, 2, 2] + 6 I[1, 2, 3, 1]\Big) - 4 I[1, 3, 2, 1] + 
       8 I[1, 3, 2, 2] 
       \nonumber\\
        &- 4 I[1, 3, 2, 3] + 8 I[1, 3, 3, 1] - 
       8 I[1, 3, 3, 2] - 4 I[1, 3, 4, 1]\Bigg) - 
    12 e_s \Big (I[0, 4, 2, 0] - 3 I[0, 4, 2, 1] + 3 I[0, 4, 2, 2] 
    \nonumber\\
        &- 
        I[0, 4, 2, 3] - 3 I[0, 4, 3, 0] + 6 I[0, 4, 3, 1] - 
        3 I[0, 4, 3, 2] + 3 I[0, 4, 4, 0] - 3 I[0, 4, 4, 1] - 
        I[0, 4, 5, 0] 
        \nonumber\\
        &+ 12 I[1, 3, 2, 1] - 12 I[1, 3, 2, 2] + 
        4 I[1, 3, 2, 3] - 24 I[1, 3, 3, 1] + 12 I[1, 3, 3, 2] + 
        12 I[1, 3, 4, 1]\Big)\Bigg\}\nonumber
 \end{align}
 
 \begin{align}
 &+\frac{ P_1}{434865438720 \pi^7}\Bigg\{8 m_c\Bigg(-12 e_c \Big (-3 I[0, 5, 1, 1] + 40 I[0, 5, 1, 2] - 
    91 I[0, 5, 1, 3] + 74 I[0, 5, 1, 4] - 20 I[0, 5, 1, 5] 
     \nonumber\\
        &- 
    11 I[0, 5, 2, 1] - 24 I[0, 5, 2, 2] + 59 I[0, 5, 2, 3] - 
    24 I[0, 5, 2, 4] + 13 I[0, 5, 3, 1] + 4 I[0, 5, 3, 2] - 
    6 I[0, 5, 3, 3] 
     \nonumber\\
        &- 5 I[0, 5, 4, 1] + 4 I[0, 5, 4, 2] + 
    10 m_c m_s \big (I[0, 4, 1, 1] - 3 I[0, 4, 1, 2] + 
       3 I[0, 4, 1, 3] - I[0, 4, 1, 4] - 2 I[0, 4, 2, 1]
        \nonumber\\
        &+ 
       4 I[0, 4, 2, 2] - 2 I[0, 4, 2, 3] + I[0, 4, 3, 1] - 
       I[0, 4, 3, 2] + 
       4 (I[1, 3, 1, 2] - 2 I[1, 3, 1, 3] + I[1, 3, 1, 4] - 
           2 I[1, 3, 2, 2]
            \nonumber\\
        &+ 2 I[1, 3, 2, 3] + 
           I[1, 3, 3, 2])\big) + 
    5 (I[1, 4, 1, 3] - 2 I[1, 4, 1, 4] + I[1, 4, 1, 5] - 
        I[1, 4, 2, 2] + I[1, 4, 2, 4] + 2 I[1, 4, 3, 2]
         \nonumber\\
        &- 
        I[1, 4, 3, 3] - I[1, 4, 4, 2])\Big) + 
 2 e_s \Big (216 I[0, 5, 1, 1] - 792 I[0, 5, 1, 2] + 
    1056 I[0, 5, 1, 3] - 600 I[0, 5, 1, 4] 
     \nonumber\\
        & + 120 I[0, 5, 1, 5] - 
    27 I[0, 5, 2, 0] - 828 I[0, 5, 2, 1] + 2250 I[0, 5, 2, 2] - 
    1860 I[0, 5, 2, 3] + 465 I[0, 5, 2, 4] 
     \nonumber\\
        &+ 81 I[0, 5, 3, 0] + 
    981 I[0, 5, 3, 1] - 1629 I[0, 5, 3, 2] + 543 I[0, 5, 3, 3] - 
    81 I[0, 5, 4, 0] - 234  I[0, 5, 4, 1] + 117 I[0, 5, 4, 2]
     \nonumber\\
        &+ 
    27 I[0, 5, 5, 0] - 27 I[0, 5, 5, 1] + 
    5 (180 I[1, 4, 1, 2] + 520 I[1, 4, 1, 3] - 500 I[1, 4, 1, 4] + 
        160 I[1, 4, 1, 5] + 27 I[1, 4, 2, 1]  
        \nonumber\\
        &+ 315 I[1, 4, 2, 2] - 
        671 I[1, 4, 2, 3] + 329 I[1, 4, 2, 4] - 
        81  I[1, 4, 3, 1] - 90 I[1, 4, 3, 2] + 151 I[1, 4, 3, 3] + 
        81 I[1, 4, 4, 1] 
         \nonumber\\
        &- 45 I[1, 4, 4, 2] - 
        27 I[1, 4, 5, 1])\Big) + 
 e_u \Big (504 I[0, 5, 1, 1] - 1788 I[0, 5, 1, 2] + 
    2324 I[0, 5, 1, 3] - 1300 I[0, 5, 1, 4] 
     \nonumber\\
        &+ 260 I[0, 5, 1, 5] + 
    9 I[0, 5, 2, 0] - 2040 I[0, 5, 2, 1] + 5112 I[0, 5, 2, 2] - 
    4108 I[0, 5, 2, 3] + 1027 I[0, 5, 2, 4] 
     \nonumber\\
        &- 27 I[0, 5, 3, 0] + 
    2577 I[0, 5, 3, 1] - 3849 I[0, 5, 3, 2] + 1283 I[0, 5, 3, 3] + 
    27 I[0, 5, 4, 0] + 66 I[0, 5, 4, 1] - 33 I[0, 5, 4, 2] 
     \nonumber\\
        &- 
    9 I[0, 5, 5, 0] + 9 I[0, 5, 5, 1] + 
    360 m_c m_s \big (2 I[0, 4, 1, 1] - 5 I[0, 4, 1, 2] + 
       4 I[0, 4, 1, 3] - I[0, 4, 1, 4] - 4 I[0, 4, 2, 1] 
        \nonumber\\
        &+ 
       6 I[0, 4, 2, 2] - 2 I[0, 4, 2, 3] + 2 I[0, 4, 3, 1] - 
       I[0, 4, 3, 2] + 
       4 (I[1, 3, 1, 2] - 2 I[1, 3, 1, 3] + I[1, 3, 1, 4] - 
           2 I[1, 3, 2, 2] 
            \nonumber\\
        &+ 2 I[1, 3, 2, 3] + 
           I[1, 3, 3, 2])\big) - 
    5 (468 I[1, 4, 1, 2] - 1412 I[1, 4, 1, 3] + 
        1420 I[1, 4, 1, 4] - 476 I[1, 4, 1, 5] + 9 I[1, 4, 2, 1] 
         \nonumber\\
        &- 
        1185 I[1, 4, 2, 2] + 2359 I[1, 4, 2, 3] - 
        1183 I[1, 4, 2, 4] - 27 I[1, 4, 3, 1] + 
        966 I[1, 4, 3, 2] - 947 I[1, 4, 3, 3] 
         \nonumber\\
        &+ 27 I[1, 4, 4, 1] - 
        249 I[1, 4, 4, 2] - 9  I[1, 4, 5, 1])\Big) + 
 e_d \Big (504 I[0, 5, 1, 1] - 1788 I[0, 5, 1, 2] + 
    2324 I[0, 5, 1, 3] 
     \nonumber\\
        &- 1300 I[0, 5, 1, 4] + 260 I[0, 5, 1, 5] + 
    9 I[0, 5, 2, 0] - 2040 I[0, 5, 2, 1] + 5112 I[0, 5, 2, 2] - 
    4108 I[0, 5, 2, 3] 
     \nonumber\\
        &+ 1027 I[0, 5, 2, 4] - 27 I[0, 5, 3, 0] + 
    2577 I[0, 5, 3, 1] - 3849 I[0, 5, 3, 2] + 1283 I[0, 5, 3, 3] + 
    27 I[0, 5, 4, 0]  \nonumber\\
        &
        + 66 I[0, 5, 4, 1] 
    - 33 I[0, 5, 4, 2] - 
    9 I[0, 5, 5, 0] + 9 I[0, 5, 5, 1] + 
    360 m_c m_s \big (2 I[0, 4, 1, 1] - 5 I[0, 4, 1, 2] + 
       4 I[0, 4, 1, 3] 
        \nonumber\\
        &- I[0, 4, 1, 4] - 4 I[0, 4, 2, 1] + 
       6 I[0, 4, 2, 2] - 2 I[0, 4, 2, 3] + 2 I[0, 4, 3, 1] - 
       I[0, 4, 3, 2] + 
       4 (I[1, 3, 1, 2] - 2 I[1, 3, 1, 3]
        \nonumber\\
        &+ I[1, 3, 1, 4] - 
           2 I[1, 3, 2, 2] + 2 I[1, 3, 2, 3] + 
           I[1, 3, 3, 2])\big) + 
    5 (468 I[1, 4, 1, 2] + 1412 I[1, 4, 1, 3] - 
        1420 I[1, 4, 1, 4] 
         \nonumber\\
        &+ 476 I[1, 4, 1, 5] - 9 I[1, 4, 2, 1] + 
        1185 I[1, 4, 2, 2] - 2359 I[1, 4, 2, 3] + 
        1183 I[1, 4, 2, 4] + 27 I[1, 4, 3, 1] 
         \nonumber\\
        &- 
        966 I[1, 4, 3, 2] + 947 I[1, 4, 3, 3] - 27 I[1, 4, 4, 1] + 
        249 I[1, 4, 4, 2] + 9 I[1, 4, 5, 1])\Big)  \Bigg)\nonumber\\
       & +
        f_{3 \gamma}m_c \Bigg(
         \Big (-45  (11 e_d + 24 e_s) I[0, 4, 3, 0] + 
   e_u (736 m_c m_s I[0, 3, 2, 0] - 459  I[0, 4, 3, 0] - 
      30 I[0, 4, 4, 0]) \nonumber\\
        &- 60 e_s I[0, 4, 4, 0]\Big) I_2[\mathcal V]
        +64 \Big ( \big (8 e_d m_c m_s I[0, 3, 2, 0] + 
      8 e_u m_c m_s I[0, 3, 2, 0] + 4 e_d m_c m_s I[0, 3, 3, 0] + 
      4 e_u m_c m_s \nonumber\\
       & \times I[0, 3, 3, 0] + e_d I[0, 4, 3, 0] + 
      e_s I[0, 4, 3, 0] + e_u I[0, 4, 3, 0] + 
      3 (3 e_d - 7 e_s + 3 e_u) I[0, 4, 4, 0]\big) I_ 5[\psi^a] 
      \nonumber\\
       &+ 
   4 m_c m_s (-4 e_d I_ 6[\psi_ {\gamma} {\nu}] I[0, 3, 2, 
         0] + (e_d + e_u) \big (-I[0, 3, 1, 0] + 7 I[0, 3, 1, 1] - 
           9 I[0, 3, 1, 2] + 3 I[0, 3, 1, 3]
           \nonumber\\
       &+ 2 I[0, 3, 2, 0] - 
           8 I[0, 3, 2, 1] + 4 I[0, 3, 2, 2] - I[0, 3, 3, 0] + 
           I[0, 3, 3, 1]\big) \psi^a[u_ 0])\Big)
           +\Big (e_s \big (-21 I[0, 4, 1, 0]
           \nonumber\\
       &+ 82 I[0, 4, 1, 1] - 
        121 I[0, 4, 1, 2] + 80 I[0, 4, 1, 3] - 20 I[0, 4, 1, 4] + 
        63 I[0, 4, 2, 0] - 185 I[0, 4, 2, 1] + 183 I[0, 4, 2, 2]
        \nonumber\\
       &- 
        61 I[0, 4, 2, 3] - 63 I[0, 4, 3, 0] + 124 I[0, 4, 3, 1] - 
        62 I[0, 4, 3, 2] + 21 I[0, 4, 4, 0] - 21 I[0, 4, 4, 1]\big) + 
     e_d \big (9 I[0, 4, 1, 0] 
     \nonumber\\
       &- 38 I[0, 4, 1, 1] + 
        59 I[0, 4, 1, 2] - 40 I[0, 4, 1, 3] + 10 I[0, 4, 1, 4] - 
        27 I[0, 4, 2, 0] + 85 I[0, 4, 2, 1] - 87 I[0, 4, 2, 2]
        \nonumber\\
       &+ 
        29 I[0, 4, 2, 3] + 27 I[0, 4, 3, 0] - 56 I[0, 4, 3, 1] + 
        28 I[0, 4, 3, 2] - 9 I[0, 4, 4, 0] + 9 I[0, 4, 4, 1]\big) + 
     e_u \big (9 I[0, 4, 1, 0]
     \nonumber\\
       &- 38 I[0, 4, 1, 1] + 
         59 I[0, 4, 1, 2] - 40 I[0, 4, 1, 3] + 10 I[0, 4, 1, 4] - 
         27 I[0, 4, 2, 0] + 85 I[0, 4, 2, 1] - 87 I[0, 4, 2, 2] 
         \nonumber\\
       &+ 
         29 I[0, 4, 2, 3] + 27 I[0, 4, 3, 0] - 56 I[0, 4, 3, 1] + 
         28 I[0, 4, 3, 2] - 9 I[0, 4, 4, 0] + 
         9 I[0, 4, 4, 1]\big)\Big) \psi^a[u_ 0] 
         \nonumber\\
       &+ 
 8 e_d m_c m_s \big (-I[0, 3, 1, 2] + I[0, 3, 1, 3] + 
     I[0, 3, 2, 2]\big) \psi_ {\gamma}^{\nu}[u_ 0]
\Bigg)
        \Bigg\}\nonumber
 \end{align}
 \begin{align}
&-\frac{m_c (3P_2-P_3)}{377487360 \pi^5} \Bigg\{ 
 (e_u+2e_d -5e_s)\Bigg (10 f_ {3\gamma} \pi^2 I_2[\mathcal V] (10 m_ 0^2 m_c I[0, 
        3, 3, 0] - 3 m_s I[0, 4, 4, 0]) + 
   3 \Big ( (-m_c I[0, 5, 3, 0] 
   \nonumber\\
       &+ 
          21 m_s I[0, 5, 4, 0]) I_ 4[\mathcal S] + 
       5 m_ 0^2 \big (3 m_s \big (-I[0, 4, 2, 0] + 
             3 I[0, 4, 2, 1] - 3 I[0, 4, 2, 2] + I[0, 4, 2, 3] + 
             3 I[0, 4, 3, 0] 
             \nonumber\\
       &- 6 I[0, 4, 3, 1] + 3 I[0, 4, 3, 2] - 
             3 I[0, 4, 4, 0] + 3 I[0, 4, 4, 1] + I[0, 4, 5, 0] - 
             12 I[1, 3, 2, 1] + 12 I[1, 3, 2, 2] - 
             4 I[1, 3, 2, 3] 
             \nonumber\\
       &+ 24 I[1, 3, 3, 1] - 
             12 I[1, 3, 3, 2] - 12 I[1, 3, 4, 1]\big) + 
          14 m_c \big (I[0, 4, 2, 0] - 3 I[0, 4, 2, 1] + 
              3 I[0, 4, 2, 2] - I[0, 4, 2, 3] 
              \nonumber\\
       &- 2 I[0, 4, 3, 0] + 
              4 I[0, 4, 3, 1] - 2 I[0, 4, 3, 2] + I[0, 4, 4, 0] - 
              I[0, 4, 4, 1] + 4 I[1, 3, 2, 1] - 8 I[1, 3, 2, 2] + 
              4 I[1, 3, 2, 3] 
              \nonumber\\
       &- 8 I[1, 3, 3, 1] + 
              8 I[1, 3, 3, 2] + 4 I[1, 3, 4, 1]\big)\big) - 
       4 \big (7 m_c \big (2 I[0, 5, 2, 1] - 5 I[0, 5, 2, 2] + 
              4 I[0, 5, 2, 3] - I[0, 5, 2, 4] 
              \nonumber\\
       &- 4 I[0, 5, 3, 1] + 
              6 I[0, 5, 3, 2] - 2 I[0, 5, 3, 3] + 2 I[0, 5, 4, 1] -
               I[0, 5, 4, 2] + 5 I[1, 4, 2, 2] - 
              10 I[1, 4, 2, 3] 
              \nonumber\\
       &+ 5 I[1, 4, 2, 4] - 
              10 I[1, 4, 3, 2] + 10 I[1, 4, 3, 3] + 
              5 I[1, 4, 4, 2]\big) + 
           3 m_s \big (I[0, 5, 2, 0] - 4 I[0, 5, 2, 1] + 
               6 I[0, 5, 2, 2] 
               \nonumber\\
       &- 4 I[0, 5, 2, 3] + I[0, 5, 2, 4] - 
               3 I[0, 5, 3, 0] + 9 I[0, 5, 3, 1] - 
               9 I[0, 5, 3, 2] + 3 I[0, 5, 3, 3] + 
               3 I[0, 5, 4, 0] - 6 I[0, 5, 4, 1]
               \nonumber\\
       &+ 
               3 I[0, 5, 4, 2] - I[0, 5, 5, 0] + I[0, 5, 5, 1] + 
               5 I[1, 4, 2, 1] - 15 I[1, 4, 2, 2] + 
               15 I[1, 4, 2, 3] - 5 I[1, 4, 2, 4] - 
               15 I[1, 4, 3, 1] 
               \nonumber\\
       &+ 30 I[1, 4, 3, 2] - 
               15 I[1, 4, 3, 3] + 15 I[1, 4, 4, 1] - 
               15 I[1, 4, 4, 2] - 
               5 I[1, 4, 5, 1]\big)\big)\Big)\Bigg)
\Bigg\}\nonumber\\
&+\frac{m_c^2 m_s}{754974720 \pi^7} \Bigg\{e_c \Big (2 I[0, 6, 1, 3] - 5 I[0, 6, 1, 4] + 4 I[0, 6, 1, 5] - 
    I[0, 6, 1, 6] - 4 I[0, 6, 2, 3] + 6 I[0, 6, 2, 4] - 
    2 I[0, 6, 2, 5] 
    \nonumber\\
       &+ 2 I[0, 6, 3, 3] - I[0, 6, 3, 4] + 
    6 I[1, 5, 1, 4] - 12 I[1, 5, 1, 5] + 6 I[1, 5, 1, 6] - 
    12 I[1, 5, 2, 4] + 12 I[1, 5, 2, 5] + 
    6 I[1, 5, 3, 4]\Big) 
    \nonumber\\
       &- (e_d + e_u) \Big (3 I[0, 6, 2, 2] - 
    7 I[0, 6, 2, 3] + 5 I[0, 6, 2, 4] - I[0, 6, 2, 5] - 
    6 I[0, 6, 3, 2] + 8 I[0, 6, 3, 3] - 2 I[0, 6, 3, 4]
    \nonumber\\
       &+ 
    3 I[0, 6, 4, 2] - I[0, 6, 4, 3] + 6 I[1, 5, 2, 3] - 
    12 I[1, 5, 2, 4] + 6 I[1, 5, 2, 5] - 12 I[1, 5, 3, 3] + 
    12 I[1, 5, 3, 4] 
    \nonumber\\
       &+ 6 I[1, 5, 4, 3]\Big) \Bigg\}\nonumber\\
       & -\frac{m_c}{21139292160 \pi^7} \Bigg\{
       e_c \Big (60 I[0, 7, 1, 3] - 210 I[0, 7, 1, 4] + 270 I[0, 7, 1, 5] - 
    150 I[0, 7, 1, 6] + 30 I[0, 7, 1, 7] 
    \nonumber\\
       &- 176 I[0, 7, 2, 3] + 
    441 I[0, 7, 2, 4] - 354 I[0, 7, 2, 5] + 89 I[0, 7, 2, 6] + 
    172 I[0, 7, 3, 3] - 260 I[0, 7, 3, 4] + 88 I[0, 7, 3, 5]
    \nonumber\\
       &- 
    56 I[0, 7, 4, 3] + 29 I[0, 7, 4, 4] + 7 I[1, 6, 2, 4] - 
    14 I[1, 6, 2, 5] + 7 I[1, 6, 2, 6] - 14 I[1, 6, 3, 4] + 
    14 I[1, 6, 3, 5]
    \nonumber\\
       &+ 7  I[1, 6, 4, 4]\Big) + 
 42 (e_d + e_s + e_u) \Big (3 I[0, 7, 2, 2] - 10 I[0, 7, 2, 3] + 
    12 I[0, 7, 2, 4] - 6 I[0, 7, 2, 5] + I[0, 7, 2, 6] 
    \nonumber\\
       &- 
    9 I[0, 7, 3, 2] + 21 I[0, 7, 3, 3] - 15 I[0, 7, 3, 4] + 
    3 I[0, 7, 3, 5] + 9 I[0, 7, 4, 2] - 12 I[0, 7, 4, 3] + 
    3 I[0, 7, 4, 4] 
    \nonumber\\
       &- 3 I[0, 7, 5, 2] + I[0, 7, 5, 3] + 
    7 I[1, 6, 2, 3] - 21 I[1, 6, 2, 4] + 21 I[1, 6, 2, 5] - 
    7 I[1, 6, 2, 6] - 21 I[1, 6, 3, 3] 
    \nonumber\\
       &+ 42 I[1, 6, 3, 4] - 
    21 I[1, 6, 3, 5] + 21 I[1, 6, 4, 3] - 21 I[1, 6, 4, 4] - 
    7 I[1, 6, 5, 3]\Big)
       \Bigg\},
 \end{align}
where 
\begin{align*}
 {M^2}= \frac{M_1^2 M_2^2}{M_1^2+M_2^2}, ~~~
 u_0= \frac{M_1^2}{M_1^2+M_2^2},
\end{align*}
%
with $ M_1^2 $ and $ M_2^2 $ being the Borel parameters in the initial and final states, respectively.
Since the same pentaquarks have existed in our initial and final states, hence we can put, M$_1^2$ = M$_2^2$= 2 M$^2$, which gives rise to $u_0 = 1/2 $.
We can interpret this result as that each quark and antiquark carry half the momentum of the photon.  Here $P_1 =\langle g_s^2 G^2\rangle$ is gluon condensate, $P_2 =\langle \bar q q \rangle$ stands for u/d-quark condensate and $P_3 =\langle \bar s s  \rangle$ denotes s-quark condensate.    The~$I[n,m,l,k]$ and $I_i[\mathcal{F}]$ functions are
defined as:
\begin{align}
 I[n,m,l,k]&= \int_{4 m_c^2}^{s_0} ds \int_{0}^1 dt \int_{0}^1 dw~ e^{-s/M^2}~
 s^n\,(s-4\,m_c^2)^m\,t^l\,w^k,\nonumber
   \end{align}
 \begin{align}
 I_1[\mathcal{F}]&=\int D_{\alpha_i} \int_0^1 dv~ \mathcal{F}(\alpha_{\bar q},\alpha_q,\alpha_g)
 \delta'(\alpha_ q +\bar v \alpha_g-u_0),\nonumber\\
  I_2[\mathcal{F}]&=\int D_{\alpha_i} \int_0^1 dv~ \mathcal{F}(\alpha_{\bar q},\alpha_q,\alpha_g)
 \delta'(\alpha_{\bar q}+ v \alpha_g-u_0),\nonumber\\
    I_3[\mathcal{F}]&=\int D_{\alpha_i} \int_0^1 dv~ \mathcal{F}(\alpha_{\bar q},\alpha_q,\alpha_g)
 \delta(\alpha_ q +\bar v \alpha_g-u_0),\nonumber\\
   I_4[\mathcal{F}]&=\int D_{\alpha_i} \int_0^1 dv~ \mathcal{F}(\alpha_{\bar q},\alpha_q,\alpha_g)
 \delta(\alpha_{\bar q}+ v \alpha_g-u_0),\nonumber\\
   I_5[\mathcal{F}]&=\int_0^1 du~ \mathcal{F}(u)\delta'(u-u_0),\nonumber\\
 I_6[\mathcal{F}]&=\int_0^1 du~ \mathcal{F}(u),
 \end{align}
 where $\mathcal{F}$ denotes the corresponding photon DAs.

 \end{widetext}

\bibliographystyle{elsarticle-num}
\bibliography{Possible_pentaquarksMM.bib}

\end{document}